\documentclass[useAMS,usenatbib]{mn2e}

\usepackage{epsfig}
\usepackage{aas_macros}
\usepackage{natbib}
\usepackage{times}

\newbox\grsign \setbox\grsign=\hbox{$>$} \newdimen\grdimen \grdimen=\ht\grsign
\newbox\labox \newbox\gabox \newbox\simpropbox \newbox\wtildebox 
\setbox\gabox=\hbox{\raise.5ex\hbox{$>$}\llap
    {\lower.5ex\hbox{$\sim$}}}\ht1=\grdimen\dp1=0pt
\setbox\labox=\hbox{\raise.5ex\hbox{$<$}\llap
    {\lower.5ex\hbox{$\sim$}}}\ht2=\grdimen\dp2=0pt
\def\ga{\mathrel{\copy\gabox}}
\def\la{\mathrel{\copy\labox}}

\newcommand{\msun}{\mbox{M$_\odot$}}

\title{What controls star formation in the central 500~pc of the Galaxy?}
\author{J.~M.~Diederik Kruijssen,$^1$\thanks{kruijssen@mpa-garching.mpg.de} Steven~N.~Longmore,$^{2,3}$ Bruce~G. Elmegreen,$^{4}$
\newauthor Norman Murray,$^{5}$ John Bally,$^{6}$ Leonardo Testi$^{3,7}$ and Robert~C.~Kennicutt, Jr.$^{8}$\\
$^{1}$Max-Planck Institut f\"{u}r Astrophysik, Karl-Schwarzschild-Stra\ss e 1, 85748 Garching, Germany\\
$^{2}$Astrophysics Research Institute, Liverpool John Moores University, IC2, Liverpool Science Park, 146 Brownlow Hill,
Liverpool L3 5RF, United Kingdom\\
$^{3}$European Southern Observatory, Karl-Schwarzschild-Stra\ss e 2, 85748, Garching, Germany\\
$^{4}$IBM T.~J. Watson Research Center, 1101 Kitchawan Road, Yorktown Heights, New York 10598, USA\\
$^{5}$Canadian Institute for Theoretical Astrophysics, 60 St. George Street, University of Toronto, Toronto, ON M5S 3H8, Canada\\
$^{6}$Center for Astrophysics and Space Astronomy, University of Colorado, Boulder, CO 80309, USA\\
$^{7}$INAF-Osservatorio Astrofisico di Arcetri, Largo E. Fermi 5, I-50125 Firenze, Italy\\
$^{8}$Institute of Astronomy, University of Cambridge, Madingley Road, Cambridge CB3 0HA, UK}

\begin{document}

\date{Accepted 2014 March 12. Received 2014 March 11; in original form 2013 March 25.}

\pagerange{\pageref{firstpage}--\pageref{lastpage}} \pubyear{2013}
\label{firstpage}

\maketitle

\begin{abstract}
The star formation rate (SFR) in the Central Molecular Zone (CMZ, i.e.~the central 500~pc) of the Milky Way is lower by a factor of $\geq10$ than expected for the substantial amount of dense gas it contains, which challenges current star formation theories. In this paper, we quantify which physical mechanisms could be responsible. On scales larger than the disc scale height, the low SFR is found to be consistent with episodic star formation due to secular instabilities or possibly variations of the gas inflow along the Galactic bar. The CMZ is marginally Toomre-stable when including gas and stars, but highly Toomre-stable when only accounting for the gas, indicating a low condensation rate of self-gravitating clouds. On small scales, we find that the SFR in the CMZ may be caused by an elevated critical density for star formation due to the high turbulent pressure. The existence of a universal density threshold for star formation is ruled out. The H{\sc i}--H$_2$ phase transition of hydrogen, the tidal field, a possible underproduction of massive stars due to a bottom-heavy initial mass function, magnetic fields, and cosmic ray or radiation pressure feedback also cannot individually explain the low SFR. We propose a self-consistent cycle of star formation in the CMZ, in which the effects of several different processes combine to inhibit star formation. The rate-limiting factor is the slow evolution of the gas towards collapse -- once star formation is initiated it proceeds at a normal rate. The ubiquity of star formation inhibitors suggests that a lowered central SFR should be a common phenomenon in other galaxies. We discuss the implications for galactic-scale star formation and supermassive black hole growth, and relate our results to the star formation conditions in other extreme environments.
\end{abstract}


\section{Introduction} \label{sec:intro}
Star formation in galactic discs is often described with a power law relation between the star formation rate (SFR) surface density $\Sigma_{\rm SFR}$ and the gas surface density $\Sigma$ \citep{kennicutt98b}:
\begin{equation}
\label{eq:sflaw}
\Sigma_{\rm SFR}=A_{\rm SK}\Sigma^N ,
\end{equation}
as was originally proposed for volume densities by \citet{schmidt59}. The typical range of power law indices is $N=1$--$2$, whether $\Sigma$ refers to dense or all gas, and across a range of spatial scales \citep{kennicutt98,kennicutt98b,bigiel08,liu11,lada12,kennicutt12}. Star-forming discs in the local Universe follow a tight, `Schmidt-Kennicutt' relation with $A_{\rm SK}=2.5\times10^{-4}$ and $N=1.4\pm0.1$ \citep{kennicutt98b}, with $\Sigma_{\rm SFR}$ in units of $\msun~{\rm yr}^{-1}~{\rm kpc}^{-2}$ and $\Sigma$ in units of $\msun~{\rm pc}^{-2}$. An exponent $N>1$ indicates that the star formation time-scale decreases with increasing surface density and could reflect some effect of self-gravity. We refer to these relations as surface density-dependent star formation relations. When using only molecular gas, $A_{\rm mol}=8\times10^{-4}$ and $N=1.0\pm0.2$ \citep{bigiel08,bigiel11}, implying that the dependence on surface area cancels and equation~(\ref{eq:sflaw}) can be written as ${\rm SFR}=A_{\rm mol}M_{\rm mol}$, where $M_{\rm mol}$ is the molecular gas mass. This is a surface density-independent star formation relation, with a constant gas depletion time-scale $t_{\rm depl}\equiv\Sigma\Sigma_{\rm SFR}^{-1}$ (or alternatively $t_{\rm depl}\equiv M_{\rm mol}/{\rm SFR}$).

Another commonly-used expression is the Silk-Elmegreen \citep{silk97,elmegreen87,elmegreen93,elmegreen97b} relation:
\begin{equation}
\label{eq:sflawomega}
\Sigma_{\rm SFR}=A_{\rm SE}\Sigma\Omega ,
\end{equation}
with $\Omega$ the angular velocity at the edge of the star-forming disc and $A_{\rm SE}=0.017$ \citep{kennicutt98b} a proportionality constant, adopting the same units as before and writing $\Omega$ in units of ${\rm Myr}^{-1}$. As for the case of $N\neq1$ in equation~(\ref{eq:sflaw}), the dependence on the angular velocity implies that the gas depletion time-scale is not constant.

These galactic-scale, global star formation relations have the advantage that they are easily evaluated observationally, but the dependence of surface densities on projection suggests that more fundamental physics drive the observed scaling relations. Recent analyses of star formation in the solar neighbourhood have been used to propose a possibly universal, local surface density threshold for star formation $\Sigma_{\rm Lada}=116~\msun~{\rm pc}^{-2}$ above which most\footnote{Except for the mass loss due to protostellar outflows \citep[e.g.][]{matzner00,nakamura07}.} gas is converted into stars on a $\sim20~{\rm Myr}$ time-scale (\citealt{gao04,heiderman10,lada10}, although see \citealt{gutermuth11} and \citealt{burkert13} for an opposing conclusion). It has been proposed that this threshold translates to a volume density threshold of $n_{\rm Lada}\sim10^4~{\rm cm}^{-3}$ and also holds on galactic scales \citep{lada12}.\footnote{We emphasise that this volume density threshold is inferred by \citet{lada10} from surface density measurements. Throughout this paper, we often choose to adopt to the volume density threshold and refer to it as the `Lada threshold', even though the original threshold refers to a surface density.} Could this volume density threshold reflect the physics of star formation across all environments, from nearby, low-mass star forming regions to high-redshift starburst galaxies?

It has been shown that $\Sigma_{\rm SFR}$ drops below the relations of equations~(\ref{eq:sflaw}) and~(\ref{eq:sflawomega}) beyond a certain galactocentric radius \citep[e.g.][]{martin01,bigiel10} -- the straightforward detection of these `cutoff' radii is well-suited for verifying the existence of star formation thresholds. However, these are also the regions of galaxies where the minority of the dense gas resides. {A strongly contrasting region within the Milky Way is the central 500~pc (the Central Molecular Zone or CMZ), in which the high gas densities and turbulent velocities are reminiscent of the extreme galactic environments found at high redshift \citep{kruijssen13}. The CMZ therefore provides an excellent opportunity to test star formation relations and theories at high spatial resolution in an extreme environment.

The observations to date conclude that the total gas mass and SFR within 200~pc of the Galactic centre lie in the ranges $3$--$7\times10^7~\msun$ and $0.04$--$0.1~\msun~{\rm yr}^{-1}$ \citep{altenhoff79,scoville87,morris96,dahmen98,ferriere07,yusefzadeh09,molinari11,immer12,longmore13}. These numerous, independent studies use different observational tracers of gas and star formation activity as well as different analysis techniques, each with different underlying assumptions.

The fact that mass and SFR {estimates} robustly converge shows that unless there is some presently-unknown, systematic error plaguing all these independent studies in the same way, the observed mass and SFR are incontrovertible to within a factor of a few. To further illustrate the agreement, we note that gas mass measurements using CO emission \citep[e.g.][]{ferriere07} rely on CO-to-H$_2$ conversion factors consistent with those derived for external galaxy centres \citep{sandstrom13}. This is confirmed in \S\ref{sec:globalobs} by comparing their gas mass measurements to those obtained using dust emission \citep{longmore13}.

The observational consensus implies that the CMZ contains 5--10~per~cent of the total star formation and 5--10~per~cent of the total molecular gas in the Galaxy. It follows trivially that the gas reservoirs in the CMZ and that in the disc of the Galaxy have the same depletion time-scales and are therefore consistent with any star formation relation in which the SFR (density) is linearly proportional to the gas mass (density). However, in all star formation relations other than the \citet{bigiel08} relation, the SFR (density) is proportional to the gas mass (density) to some power (e.g.~$N=1.4$ in equation~\ref{eq:sflaw}) or depends on a second parameter (e.g.~$\Omega$ in equation~\ref{eq:sflawomega}). As a result, the gas depletion time-scale in these relations is not constant but depends on the surface density or angular velocity. Therefore, the only way that the gas in the CMZ and the Galactic disc could simultaneously satisfy these relations is if they have the same gas surface density and angular velocity.

In \citet{longmore13}, we have shown that the CMZ contains $\sim80$~per~cent of the NH$_3(1,1)$ integrated intensity in the Galaxy \citep{longmore13}, reflecting an overwhelming abundance of dense gas ($\Sigma>10^2~\msun~{\rm pc}^{-2}$ and $n>{\rm~a~few}\times10^3~{\rm cm}^{-3}$). The surface and volume densities of gas in the CMZ are found to be on average 1--2 orders of magnitude larger than that in the disc -- \emph{it is therefore impossible to reconcile both the CMZ and the disc with volumetric and surface density star formation relations predicting that a given mass of gas will form stars more rapidly if the surface or volume density is higher} \citep[e.g.][]{kennicutt98b,krumholz05,padoan11,krumholz12a}.} Indeed, these relations over-predict the measured star formation rate by factors of 10 to 100 \citep{longmore13}. The low SFR in the CMZ is particularly striking because its gas surface density is similar to that observed in high-redshift regular disc and starburst galaxies, some of which seem to have a factor of ten \textit{higher} SFRs than predicted by typical star formation relations \citep{daddi10b,genzel10}. If the gas depletion time-scale depends on the density, something is required to slow the rate of star formation in the CMZ compared to the rest of the Milky Way and other galaxies.

The SFR in the CMZ is also inconsistent with the \citet{lada12} star formation relation. Because most gas in the CMZ is residing at {surface} densities larger than the Lada surface density threshold (the same holds for the implied volume densities), a gas consumption time-scale of 20~Myr gives a SFR of $\sim2~\msun~{\rm yr}^{-1}$, which is 1--2 orders of magnitude higher than the measured SFR of $0.04$--$0.1~\msun~{\rm yr}^{-1}$ \citep{yusefzadeh09,immer12,longmore13}.

In nearby disc galaxies, a simple proportionality of the SFR to the molecular mass is commonplace \citep{bigiel08,bigiel11}. Despite the similarities between the CMZ and high-redshift galaxies \citep{kruijssen13}, which do form stars at or above the rate predicted by surface density-dependent star formation relations, the CMZ is consistent with an extrapolation of the molecular star formation relation with a constant H$_2$ depletion time-scale (i.e.~$N=1$ in equation~\ref{eq:sflaw}, also see below). This is surprising -- self-gravity implies that dynamical evolution proceeds faster when the surface density is higher (i.e.~$N>1$). If gravity is an important driver of star formation in the CMZ and galaxy discs, then {\it a constant molecular gas depletion time-scale requires that some resistance to the gravitational collapse towards stars must increase to offset the effect of self-gravity as the surface density goes up}. Due to its extreme characteristics, the CMZ provides an excellent opportunity for understanding the underlying physics.

\begin{figure*}
\center\resizebox{\hsize}{!}{\includegraphics{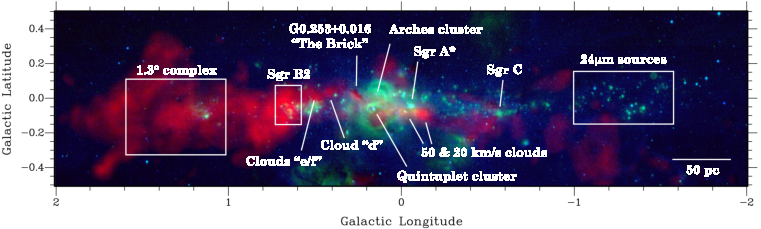}}\\
\caption[]{\label{fig:img}
      {Three-colour composite of the CMZ, with in red the HOPS NH$_3(1,1)$ emission \citep{walsh11,purcell12} to indicate the gas with a volume density above a few times $10^3~{\rm cm}^{-3}$, in green the MSX 21.3$\mu{\rm m}$ image \citep{egan98,price01}, and in blue the MSX 8.28$\mu{\rm m}$ image. The MSX data shows PAH emission (mostly tracing cloud edges), young stellar objects, and evolved stars. The labels indicate several key objects and regions.}
                 }
\end{figure*}
In this paper, we take the point of view that the star formation relations of equation~(\ref{eq:sflaw}) (with $N>1$) and equation~(\ref{eq:sflawomega}) do not seem to apply to the CMZ of the Milky Way. To understand this difference better, we evaluate the global and local processes that affect the rate of star formation in the central region of the Milky Way. Thanks to the strongly contrasting environments of galaxy centres and galaxy discs, galaxy centres provide a unique and independent way to study the universality of star formation relations. We exploit this contrast to discuss the implications of the lack of star formation in the CMZ for star formation relations with $N>1$. The paper is concluded by sketching a possible picture of how local and global star formation criteria connect, and we propose observational and numerical tests through which the different components of this picture can be addressed in more detail.

Throughout the paper, we adopt a mean molecular weight of $\mu=2.3$, implying a mean particle mass of $\mu m_{\rm H}=3.9\times10^{-24}~{\rm g}$.

\section{Observational constraints from the Central Molecular Zone} \label{sec:obs}
\citet{longmore13} present the observational constraints for the suppression of star formation in the CMZ, which are summarized here. A division is made between {\it global} and {\it local} physics, where `global' refers to spatial scales larger than the disc scale height ($\Delta R>h$), on which the ISM properties are set by galactic structure, and `local' refers to spatial scales smaller than the disc scale height ($\Delta R<h$), on which the ISM properties are set by cloud-scale physics. For instance, the formation of giant molecular clouds (GMCs) proceeds on global scales, whereas processes internal to the GMCs, such as stellar feedback, the turbulent cascade, or magnetic fields constitute local physics, which determine the fraction of the GMC mass that proceeds to star formation. The scale height of the gas in the CMZ is 10--50~pc (see Table~\ref{tab:prop}).

\subsection{Global constraints} \label{sec:globalobs}
Figure~\ref{fig:img} shows a three-colour composite of the central 4$^\circ$ of the Milky Way, corresponding to a spatial scale of $\sim600~{\rm pc}$ at the distance of the Galactic centre \citep[we adopt $8.5$~kpc, which is consistent with][]{reid09}. We combine the dense gas emission (NH$_3(1,1)$, in red) with infrared imaging (green and blue) to highlight the gas close to (or above) the \citet{lada10} volume density threshold, young stellar objects, and evolved stars. The legend indicates the objects in the CMZ that are most relevant to this paper. In particular, the 1.3$^\circ$ complex and the 100-pc, twisted ring of dense gas clouds between Sgr~C and Sgr~B2 \citep{molinari11} will be scrutinised in detail. The G0.253+0.016 cloud (the `Brick') was recently identified as a possible progenitor of a young massive cluster \citep{longmore12}, and we will use it as a template for the massive and dense clouds that populate the CMZ.

In Table~\ref{tab:prop}, we list the derived properties of the gas in the CMZ, as well as the typical characteristics of galaxies as defined by existing, empirical star formation relations. The columns indicate (1) the object ID, (2) the CO--H$_2$ conversion factor used to derive the gas mass (numbers in parentheses indicate that the gas mass is measured by other means), (3) the gas surface density, (4) the epicyclic frequency, (5) the velocity dispersion, (6) the scale height, (7) the \citet{toomre64} $Q$ stability parameter of the gas disc (see \S\ref{sec:surfacethr}), {(8) the stellar surface density, (9) the Toomre $Q$ parameter of the entire disc}, (10) the observed star formation rate surface density, and (11) the gas depletion time-scale. The first three rows list the properties of the CMZ, where we distinguish two components (the 100-pc ring described by \citealt{molinari11} and the $1.3^\circ$ complex), as well as its properties {smeared out} over a radius of 230~pc (corresponding to 1.55$^\circ$, i.e.~just beyond the 1.3$^\circ$ complex). This division into sub-regions is made because the 100-pc ring and the $1.3^\circ$ complex contain most of the dense (and thereby total) gas mass within $|l|\la1.5^\circ$.

\begin{table*}
 \centering
  \begin{minipage}{174.5mm}
  \caption{Properties of the considered regions and galaxies.}\label{tab:prop}
  \begin{tabular}{@{}l c c c c c c c c c c@{}}
  \hline 
  Object ID & $X_{{\rm CO},20}$ & $\Sigma_{2}$ & $\kappa$ & $\sigma$ & $h_2$ & $Q_{\rm gas}$ & $\Sigma_{\star,2}$ & $Q_{\rm tot}$ & $\Sigma_{\rm SFR}$ & $t_{\rm depl}$\\
  (1) & (2) & (3) & (4) & (5) & (6) & (7) & (8) & (9) & (10) & (11)\\
 \hline
CMZ 230~pc-integrated & $0.5$ & $1.2$ & $0.75$--$1.2$ & $10$--$20$ & $0.5$ & $(3.8$--$15)$ & $38$ & $(1.0$--$2.0)$ & $0.20$ & $(0.6)$ \\
CMZ 1.3$^\circ$ complex & $(0.5)$ & $2.0$ & $0.88$--$1.5$ & $10$--$15$ & $0.5$ & $(2.7$--$6.8)$ & $20$ & $(1.5$--$3.0)$ & $0.13$ & $(1.5)$ \\
CMZ 100-pc ring & $(0.7)$ & $30$ & $1.6$--$3.3$ & $10$--$20$ & $0.1$ & $(0.40$--$1.7)$ & $29$ & $(0.37$--$1.4)$ & $3.0$ & $(1.0)$ \\
\citet{kennicutt98b} low $\Sigma$ disc & $2.8$ & $0.05$ & $(0.04)$ & $(2.6)$ & $(0.5)$ & $1.5$ & - & - & $0.0024$ & $(2.1)$ \\
\citet{kennicutt98b} high $\Sigma$ disc & $2.8$ & $0.20$ & $(0.07)$ & $(5.9)$ & $(0.6)$ & $1.5$ & - & - & $0.017$ & $(1.2)$ \\
\citet{daddi10b} low $\Sigma$ starburst & $0.4$ & $3.0$ & $(0.34)$ & $(6.1)$ & $(0.05)$ & $1.5$ & - & - & $3.9$ & $(0.08)$ \\
\citet{daddi10b} high $\Sigma$ starburst & $0.4$ & $10^2$ & $(0.80)$ & $(86)$ & $(0.3)$ & $1.5$ & - & - & $560$ & $(0.02)$ \\
\citet{daddi10b} low $\Sigma$ BzK & $1.8$ & $2.0$ & $(0.05)$ & $(55)$ & $(6)$ & $1.5$ & - & - & $0.28$ & $(0.7)$ \\
\citet{daddi10b} high $\Sigma$ BzK & $1.8$ & $10$ & $(0.07)$ & $(200)$ & $(10)$ & $1.5$ & - & - & $2.7$ & $(0.4)$ \\
\hline
\end{tabular}\\
$X_{{\rm CO},20}\equiv X_{\rm CO}/10^{20}~({\rm K}~{\rm km}~{\rm s}^{-1}~{\rm cm}^2)^{-1}\approx0.47\alpha_{\rm CO}/\msun~({\rm K}~{\rm km}~{\rm s}^{-1}~{\rm pc}^2)^{-1}$ is the CO-to-H$_2$ conversion factor, $\Sigma_{2}\equiv\Sigma/10^2~\msun~{\rm pc}^{-2}$ is the gas surface density, $\kappa$ is the epicyclic frequency in units of ${\rm Myr}^{-1}$, $\sigma$ is the velocity dispersion in units of ${\rm km}~{\rm s}^{-1}$, $h_2\equiv h/10^2~{\rm pc}$ is the scale height, $Q_{\rm gas}$ is the \citet{toomre64} disc stability parameter {when only including the self-gravity of the gas, $\Sigma_{\star,2}\equiv\Sigma_\star/10^2~\msun~{\rm pc}^{-2}$ is the stellar surface density, $Q_{\rm tot}$ includes the self-gravity of gas {\it and} stars}, $\Sigma_{\rm SFR}$ is the star formation rate surface density in units of $\msun~{\rm yr}^{-1}~{\rm kpc}^{-2}$, and $t_{\rm depl}\equiv\Sigma\Sigma_{\rm SFR}^{-1}/{\rm Gyr}$ is the gas depletion time. Values in parentheses are implied by the other numbers (see text).
\end{minipage}
\end{table*}
The gas surface density within 230~pc is calculated by using the total gas mass from \citet{ferriere07}. The projected, face-on gas surface densities of the 100-pc ring and the $1.3^\circ$ complex are derived by calculating their gas masses from the HiGAL data \citep{molinari10} following the analysis in \citet{longmore13},\footnote{We have used dust emission to estimate these surface densities. This method is insensitive to the conversion of CO intensity to H$_2$ surface density (the ratio of which is represented by the parameter $X_{\rm CO}$) and its associated uncertainty \citep[cf.][and references therein]{sandstrom13}. However, the derived surface density does depend on the assumed gas-to-dust ratio, which may be lower in the CMZ than in the Galactic disc due to the higher metallicity (see \citealt{longmore13} for a discussion). We compare our gas mass measurements using dust emission to those from \citealt{ferriere07} using CO emission and list the resulting values of $X_{\rm CO}$ in Table~\ref{tab:prop}. We note that our derived values coincide with those derived by \citealt{sandstrom13} for external galaxy centres, i.e.~$X_{\rm CO}=0.7^{+1.7}_{-0.4}\times10^{20}$~$({\rm K}~{\rm km}~{\rm s}^{-1}~{\rm cm}^2)^{-1}$.} and adopting a certain 3D geometry. For the 100-pc ring ($M\sim10^7~\msun$), the face-on surface density is obtained by using the 3D geometry from \citet[assuming a ring thickness of 10~pc]{molinari11}. Similarly, the face-on projected surface density of the $1.3^\circ$ complex ($M\sim3\times10^6~\msun$) is obtained by assuming that in the Galactic plane it traces an ellipse with semi-major and minor axes of 85~pc and 50~pc, respectively \citep{sawada04}.

The epicyclic frequency is calculated from the rotation curve of \citet[Figure~14]{launhardt02}.\footnote{\label{footnote:uncertain}{Note that this gives circular velocities of $140$--$200~{\rm km~s}^{-1}$ for the three regions listed in Table~\ref{tab:prop}, whereas the peak line-of-sight velocity of the 100-pc ring as measured from the HOPS NH$_3$(1,1) emission \citep{walsh11,purcell12} is $80~{\rm km}~{\rm s}^{-1}$ at the location of Sgr~B2. Such a low line-of-sight velocity may be caused by the possible eccentric orbit of the ring \citep{molinari11}, which is thought to lie under such an angle that the edge of the ring lies close to apocentre as seen from Earth. In this scenario, the measured line-of-sight velocity should be lower than the local circular velocity at the position of Sgr~B2, a difference that would be amplified further by possible projection effects. An alternative explanation is that the 100-pc ring extends to higher longitudes than stated in \citet{molinari11}, in which case the circular velocity would exceed the observed line-of-sight velocities at the presumed tangent points due to projection alone (Bally et al. in prep.). This picture is consistent with the proper motion of Sgr~B2, which is $\sim90~{\rm km}~{\rm s}^{-1}$ \citep{reid09}. Using the measured velocities instead of the \citet{launhardt02} rotation curve gives a factor of $\sim1.6$ lower epicyclic frequencies. Both extremes are used to calculate the possible range of $\kappa$ listed in Table~\ref{tab:prop}, and hence also contribute to the range of $Q_{\rm gas}$ and $Q_{\rm tot}$.}} To determine the scale height of the dense gas in both the 100-pc ring and the $1.3^\circ$ complex, as well as its linewidth (i.e.~the full width at half-maximum) $\Delta V$ and hence the velocity dispersion $\sigma\equiv 2\sqrt{2\ln{2}}\Delta V$, we have used the HOPS NH$_3$(1,1) emission \citep{walsh11,purcell12}. The value ranges in column~5 of Table~\ref{tab:prop} represent the maximum and minimum measured velocity dispersions across each region. We calculate $Q_{\rm gas}$ from the preceding columns as in equation~(\ref{eq:Q}) in \S\ref{sec:surfacethr} below.

As can be seen by comparing columns~3 and~8 of Table~\ref{tab:prop}, the stellar component dominates the gravitational potential in the 230~pc-integrated CMZ and the 1.3$^\circ$ complex. The stellar surface density is obtained from the adopted rotation curve \citep{launhardt02} as follows. {We first derive a spherically symmetric volume density profile from the enclosed mass profile. The stellar surface density is then calculated by only including the stellar mass in a slab of thickness of $2h$. This ensures that the stellar surface density is obtained for the same spatial volume as for the gas,} and implies that the total stellar surface density of the bulge at the same galactocentric radii is higher than listed in Table~\ref{tab:prop}.

Column~9 gives Toomre $Q_{\rm tot}$, which is corrected for the presence of stars as in equation~(\ref{eq:Qtot}) in \S\ref{sec:surfacethr} below, using a stellar velocity dispersion of $\sigma_\star\sim100~{\rm km~s}^{-1}$ \citep{dezeeuw93}. Note that if we assume that the 1.3$^\circ$ complex is approximately spherically symmetric, its virial parameter \citep{bertoldi92} $\alpha\sim2$ indicates it is roughly in equilibrium, but only when including the stellar gravity. If the stars would be absent, the cloud would be highly unbound and hence stable against gravitational collapse. A similar effect of the stellar potential is seen when comparing Toomre $Q$ of columns~7 and~9 for the 230~pc-integrated CMZ. By contrast, the $Q$ parameter of the 100-pc ring is hardly affected by the presence of stars.

The SFR densities $\Sigma_{\rm SFR}$ in column~10 are derived using Table~2 of \citet{longmore13}, which lists the SFR for certain parts of the CMZ. These SFRs are derived from free-free emission by expressing the SFR in terms of the measured ionising luminosity \citep{mckee97,murray10b,lee12}. The resulting SFRs are consistent with those measured by other methods, such as young stellar object (YSO) counts \citep{yusefzadeh09,immer12}. For details on the derivation, we refer to the discussion in \S5 of \citet{murray10b} and \citet{longmore13}.

The gas depletion time in column~11 follows from $\Sigma$ and $\Sigma_{\rm SFR}$ as $t_{\rm depl}\equiv\Sigma\Sigma_{\rm SFR}^{-1}$. While the listed values reflect the depletion times of all gas, i.e.~including both H{\sc i} and H$_2$, it is important to note that at the high surface densities of the CMZ, most of the gas is in molecular form. Only for the disc sample from \citet[see below]{kennicutt98b} this does not hold, in which case the molecular depletion times are shorter than those for all gas listed in Table~\ref{tab:prop}. Also note that while recent observations are beginning to quantify galactic (molecular) mass outflow rates in detail \citep{crocker12b,bolatto13}, this definition of the depletion time only includes star formation and does not account for the removal of gas mass by galactic winds.

The fourth and fifth rows of Table~\ref{tab:prop} span the star formation relation of equations~(\ref{eq:sflaw}--\ref{eq:sflawomega}) for the disc galaxy sample used in \citet{kennicutt98b}. The bottom rows span the high-SFR\footnote{This extra sequence of star-forming galaxies contains some galaxies from the nearby starburst sample of \citet{kennicutt98b}, but has an elevated SFR with respect to that paper because a different value of $X_{\rm CO}$ is assumed \citep{daddi10b}. We emphasise that inferring the H$_2$ surface density using CO emission is an indirect method and hence may introduce substantial uncertainty. However, recent results from the KINGFISH survey of nearby galaxies suggest that $X_{\rm CO}$ weakly decreases with star formation rate surface density \citep[e.g.][]{sandstrom13}, which is at least qualitatively consistent with \citet{daddi10b}.} and low-SFR sequences of \citet[also see \citealt{genzel10}]{daddi10b}. At a given surface density, the typical epicyclic frequencies of the \citet{kennicutt98b} galaxies are obtained by combining equations~(\ref{eq:sflaw}) and~(\ref{eq:sflawomega}), whereas for \citet{daddi10b} they follow from their equations (2) and (3). The extragalactic velocity dispersions are obtained by assuming $Q=1.5$ and using the definition of $Q$ (see \S\ref{sec:surfacethr}). The characteristic disc scale heights are added to distinguish between the aforementioned `global' and `local' regimes, and are calculated by assuming an equilibrium disc including a factor-of-two increase of the disc self-gravity due to the presence of stars \citep[cf.][]{elmegreen89,martin01}.

\begin{figure}
\center\resizebox{\hsize}{!}{\includegraphics{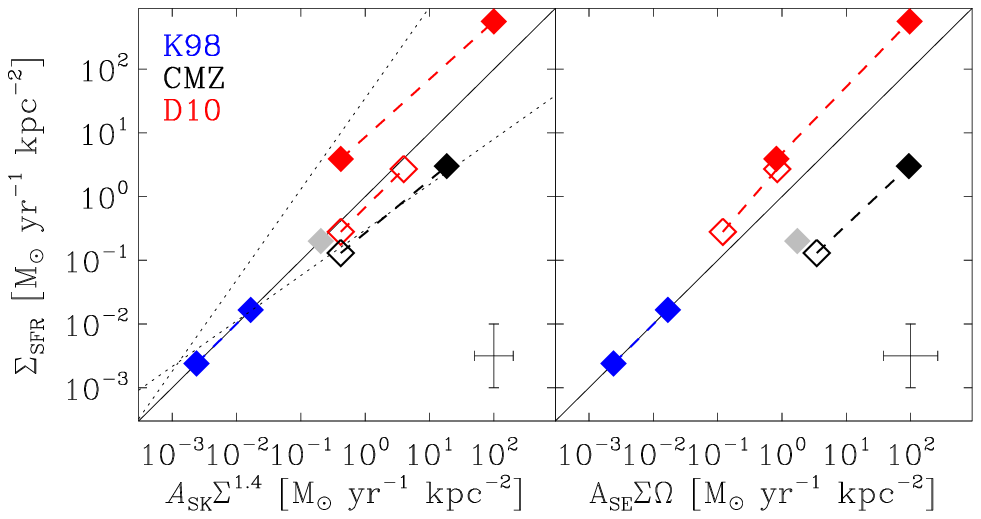}}\\
\caption[]{\label{fig:kslaw}
      Observed star formation rate surface density as a function of the global star formation relations from \citet{kennicutt98b}. {\it Left}: using the Schmidt-Kennicutt relation from equation~(\ref{eq:sflaw}). {\it Right}: using the Silk-Elmegreen relation from equation~(\ref{eq:sflawomega}). The blue symbols span the sequence of nearby disc galaxies from \citet{kennicutt98b}, the red symbols span the star-forming galaxies from \citet{daddi10b}, the open and closed, black symbols indicate the 1.3$^\circ$ complex and the 100-pc ring, respectively (see Table~\ref{tab:prop}), and the grey symbol denotes the spatially integrated CMZ. The solid lines indicate the 1:1 agreement. The dotted lines in the left panel are included for reference and represent $\Sigma_{\rm SFR}=A_{\rm mol}\Sigma$ (bottom, using $A_{\rm mol}=8\times10^{-4}$ as in \citealt{bigiel08}) and $\Sigma_{\rm SFR}=A_{\rm SK}\Sigma^2$ (top).
                 }
\end{figure}
We compare the data from Table~\ref{tab:prop} to the global star formation relations of equations~(\ref{eq:sflaw}) and~(\ref{eq:sflawomega}) in Figure~\ref{fig:kslaw}. The spatially resolved elements of the CMZ (open and closed, black symbols) are forming stars at a rate that is a factor of 3--20 (i.e.~typically an order of magnitude) below either relation. By contrast, the Schmidt-Kennicutt relation does describe the CMZ well when spatially smoothing it over a 230~pc radius (the closed, grey symbol), whereas smoothing hardly affects the agreement with the Silk-Elmegreen relation.

The contrasting agreement of the spatially averaged and resolved SFRs with the Schmidt-Kennicutt relation may simply illustrate that global star formation relations fail at spatial scales smaller than $\sim500$~pc \citep[e.g.][]{bigiel08,leroy13}. We have recently shown that this breakdown arises from the incomplete statistical sampling of the evolutionary time-sequence of star formation on small spatial scales \citep{kruijssen14}. In other words, a sufficiently large number of independent star-forming regions is required to retrieve the global star formation relation. However, the critical size-scale for this breakdown is not universal and depends on the characteristic size and time-scales involved in the physics of cloud-scale star formation. In \citet{kruijssen14}, we find that for the conditions of the CMZ, global star formation relations are expected to break down below 80~pc. Assuming that there is no additional physical process that synchronises the evolutionary stages of the star-forming regions in the CMZ (see for instance \S\ref{sec:episodic}), this implies that both the 100-pc ring and the 230-pc averaged CMZ should be consistent with global star formation relations -- but only the latter actually is.

Which representation of the CMZ is then physically appropriate? Should we smoothen the structure in the CMZ to large scales as is done for galaxy discs? The smoothing of surface densities is justified in galaxy discs, because the lifetime of substructure is typically of the order of (or shorter than) an orbital timescale \citep{dobbs14}, and accounting for substructure would therefore only introduce spurious stochasticity \citep[e.g.][]{schruba10}. However, the nuclear rings that appear in numerical simulations of barred galaxies are persistent over many dynamical times \citep[e.g.][]{piner95,kim12c}, and hence it seems physically incorrect to smear out the 100-pc ring and any other, possibly persistent structure in the CMZ to a much larger scale when comparing to global star formation relations. These structures can be present over sufficiently long time-scales to consistently affect the star formation process and hence should be accounted for when describing the star formation (or lack thereof) in the CMZ. Therefore, the physically appropriate representation of the CMZ in Figure~\ref{fig:kslaw} is given by the open and closed, black symbols.

Also visible in the left-hand panel of Figure~\ref{fig:kslaw} is that the CMZ (of which nearly all of the gas is molecular) agrees well with the \citet{bigiel08} star formation relation between the SFR and the molecular gas mass (shown as the lower dotted line), which was derived for nearby disc galaxies with surface densities $3<\Sigma_{\rm mol}/\msun~{\rm pc}^{-2}<50$. The CMZ extends this range by roughly two orders of magnitude. \citet{bigiel08,bigiel11} show that the gas depletion time of the galaxies in their sample is $t_{\rm depl}=1$--2~Gyr, which is indeed similar to the time-scales listed for the CMZ in Table~\ref{tab:prop}. In the framework of this relation, the CMZ is somehow the norm while both the \citet{daddi10b} samples of BzK \citep{tacconi10,daddi10} and starburst/submillimeter galaxies \citep{kennicutt98b,bouche07,bothwell10} are the exception. The Bigiel relation is the only global star formation relation that fits the CMZ -- the region remains anomalous in the context of equations~(\ref{eq:sflaw}) and~(\ref{eq:sflawomega}), or the \citet{lada12} relation (see \S\ref{sec:localobs}). As will be shown in \S\ref{sec:thresholds}, there are several possible reasons why the CMZ is peculiar.

\subsection{Local constraints} \label{sec:localobs}
Turning to local physics, in \citet[Figure~4]{longmore13} we have shown that 70--90 per~cent of the gas in the CMZ resides at surface densities above the Lada surface density threshold, and hence should be vigorously forming stars if the threshold is universal. Throughout this paper, we use the corresponding volume density threshold of $n_{\rm Lada}\sim10^4~{\rm cm}^{-3}$ \citep{lada10}. While this threshold may apply for the low ($\sim10^2~{\rm cm}^{-3}$) densities of GMCs in the solar neighbourhood, the mean volume density of the gas in the CMZ is $n_0\sim2\times10^4~{\rm cm}^{-3}$ \citep{longmore13}, i.e.~more than two orders of magnitudes higher, and it hosts GMCs with typical densities up to $\sim10^5~{\rm cm}^{-3}$. As mentioned in \S\ref{sec:intro}, the threshold for star formation and depletion time-scale from \citet{lada12} would imply a SFR that is 1--2 orders of magnitude higher than the measured value \citep{longmore13}.

If we assume that the SFR is driven by self-gravity, we can use the observed SFR to derive the fraction of the gas mass residing above some volume density threshold for star formation. Using Eqs.~(19)--(21) of \citet{krumholz05}, we find that the mass fraction of gas above the threshold density is
\begin{equation}
\label{eq:fsfr}
f_{\rm th}=\frac{\phi_t t_{\rm ff}}{\epsilon_{\rm core}t_{\rm depl}} ,
\end{equation}
where $\phi_t=1.91$ is a constant that indicates the ratio between the star formation time-scale and the free-fall time, and $\epsilon_{\rm core}=0.5$ is the maximum star formation efficiency in protostellar cores. Equation~(\ref{eq:fsfr}) assumes that the replenishment of gravitationally unstable gas from the cloud scale occurs on the mean free-fall time of the gas\footnote{This assumption is consistent with the observational definition of a threshold density, as can be illustrated using the Lada threshold in the solar neighbourhood. \citet{lada10} find a depletion time-scale for the dense gas of $t_{\rm depl}=20~{\rm Myr}$, with $f_{\rm th}=1$ by definition because only gas above the threshold is considered. Substitution into equation~(\ref{eq:fsfr}) then gives $t_{\rm ff}\sim5~{\rm Myr}$, which is indeed the free-fall time on the cloud scale at the densities characteristic of solar neighbourhood clouds ($n\sim50~{\rm cm}^{-3}$).} \citep{krumholz05}, i.e.~$t_{\rm ff}\sim0.25~{\rm Myr}$ for $n_0\sim2\times10^4~{\rm cm}^{-3}$. In combination with the observed depletion time-scale of molecular hydrogen in the CMZ ($t_{\rm depl}\sim1~{\rm Gyr}$, see the last column of Table~\ref{tab:prop}), this implies that only a fraction $f_{\rm th}\sim0.001$ of the gas mass in the CMZ {\it should} be above some volume density threshold.

\begin{figure}
\center\resizebox{7.6cm}{!}{\includegraphics{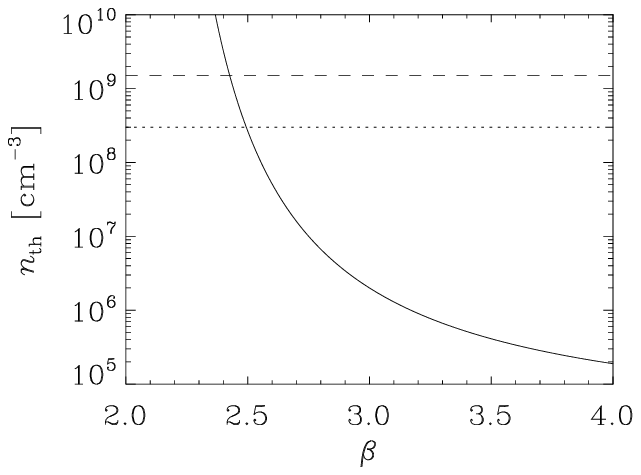}}\\
\caption[]{\label{fig:nth}
      Volume density threshold for star formation implied by the lack of star formation in the Central Molecular Zone of the Milky Way as a function of the assumed power law slope of the volume density PDF at densities $n>n_{\rm Lada}$ (solid line). The dashed line indicates the required threshold density for a log-normal PDF \citep{padoan97} with Mach number ${\cal M}=30$ and mean density $n_0=2\times10^4~{\rm cm}^{-3}$, appropriate for the CMZ (see text). The dotted line indicates the same, but accounts for the effect on the density PDF of a magnetic field with strength $B\sim100\mu{\rm G}$ at a temperature of $T=65$~K.
                 }
\end{figure}
The above mass fraction can be related to a density threshold by integration of the density probability distribution function (PDF) ${\rm d}p/{\rm d}n$ by writing
\begin{equation}
\label{eq:pdf}
f_{\rm th}=\frac{\int_{n_{\rm th}}^\infty n ({\rm d}p/{\rm d}n){\rm d}n}{\int_0^\infty n ({\rm d}p/{\rm d}n){\rm d}n} .
\end{equation}
The volume density PDF of a turbulent interstellar medium (ISM) is often represented by a log-normal function, of which the width and dispersion are set by the Mach number \citep[e.g.][]{vazquez94,padoan97}.\footnote{Note that the exact dependence of the width on the Mach number depends on whether the turbulence forcing is compressive or solenoidal \citep{federrath10}. In this paper, we assume the commonly-used combination of both \citep{padoan97}.} However, the high-density, self-gravitating tail can also (at least locally) be approximated by a power law, i.e.~${\rm d}p/{\rm d}n\propto n^{-\beta}$ \citep[e.g.][]{klessen00b,kritsuk11,elmegreen11,hill12}. This can be used to estimate the required value of $n_{\rm th}$ in the CMZ. As mentioned above, the fraction of gas that is used to form stars is $f_{\rm th}\sim0.001$. For a power law density PDF in the range $[n_{\rm min},\infty\rangle$, the lower limit is defined by the mean density as $n_{\rm min}=n_0(\beta-2)/(\beta-1)$ and equation~(\ref{eq:pdf}) yields
\begin{equation}
\label{eq:nthpl}
n_{\rm th}=f_{\rm th}^{1/(2-\beta)}n_0(\beta-2)/(\beta-1) .
\end{equation}
This relation is shown in Figure~\ref{fig:nth}, which gives $3\times10^8>n_{\rm th}/{\rm cm}^{-3}>10^5$ for exponents $2.5<\beta<4$. The density PDF of the gas in the CMZ is unknown, but the typical high-density slope due to self-gravity in the numerical simulations of \citet{kritsuk11} is $\beta=2.5$--$2.75$, which suggests $n_{\rm th}=10^7$--$3\times10^8~{\rm cm}^{-3}$. Note that because a power-law tail implies some effect of self-gravity, this functional form should fail to describe the detailed shape of the PDF in the low density-regime where self-gravity is not important. This needs to be kept in mind when choosing a roughly representative value of $\beta$.

For a log-normal PDF with a Mach number of ${\cal M}\sim30$ (cf.~Table~\ref{tab:prop}, with a temperature of $T=65$~K as in \citealt{ao13} -- also see \citealt{morris83} and \citealt{mills13}) and a mean density of $n_0=2\times10^4~{\rm cm}^{-3}$, the required threshold density is at the high end of the above range, with $n_{\rm th}\sim10^9~{\rm cm}^{-3}$.\footnote{If we assume that the gas is replenished on the local free-fall time instead of that on the cloud-scale \citep{hennebelle11,federrath12}, then the free-fall time from equation~(\ref{eq:fsfr}) must be included in the integral over the PDF (as a factor $1/t_{\rm ff}$ in the numerator of equation~\ref{eq:pdf}) to reflect the high-density, collapsing gas. This yields shorter free-fall times and hence the low SFR then requires a higher density threshold of $n_{\rm th}\sim10^{11}~{\rm cm}^{-3}$.} However, this does not account for the influence of the strong magnetic field \citep[$B\sim100\mu{\rm G}$,][]{crocker10} near the Galactic Centre. Thermal-to-magnetic pressure ratios of $2c_{\rm s}^2/v_A^2<1$ (where $c_{\rm s}$ is the sound speed and $v_A$ is the Alfv\'{e}n velocity) decrease the dispersion of the log-normal density PDF. Again assuming $T=65$~K, we find $2c_{\rm s}^2/v_A^2=0.31$ for the CMZ. If we modify the density PDF accordingly (cf.~\citealt{padoan11,molina12}, adopting $B\propto n^{1/2}$ as in \citealt{padoan99}), the threshold density required by the observed SFR becomes $n_{\rm th}\sim3\times10^8~{\rm cm}^{-3}$. We conclude that the CMZ firmly rules out a density threshold at $n=10^4~{\rm cm}^{-3}$, and adopt a lower limit of $n_{\rm th}=10^7~{\rm cm}^{-3}$. This is still exceptionally high in comparison to the threshold densities of nearby disc galaxies, and is only known to be reached at any given time by a substantial fraction of the gas mass in dense, rapidly star-forming galaxies at high redshift \citep[e.g.][]{swinbank11}.

\section{Possible mechanisms for inhibiting star formation}  \label{sec:thresholds}
In this section, we summarize and quantify which physical mechanisms may limit the SFR in the central regions of galaxies {with respect to the SFRs predicted by density-dependent star formation relations}. In \S\ref{sec:impl}, we propose ways of distinguishing their relative importance observationally and numerically. This section is separated into global and local star formation inhibitors. We discuss in \S\ref{sec:disc} how these can be connected.

\subsection{Globally inhibited star formation} \label{sec:global}
We first discuss the possible mechanisms that may limit star formation in the CMZ on spatial scales larger than the disc scale height ($\Delta R>h$). An apparent lack of star formation may be caused by a stability of the gas disc against gravitational collapse, or by global dynamical processes that may synchronise the evolutionary stages of CMZ clouds or limit the cloud-scale duration of star formation.

\subsubsection{Disc stability} \label{sec:surfacethr}
A recurring question regarding galactic-scale star formation relations has been in which part of the parameter space they apply -- is there a threshold (surface) density below which the SFR is negligible? And if so, by which factor is the SFR reduced? The existence of a threshold is suggested by the sharply truncated H{\sc ii} discs in galaxies, which indicate that beyond a certain galactocentric radius $\Sigma_{\rm SFR}$ falls off more rapidly than suggested by equations~(\ref{eq:sflaw}) and~(\ref{eq:sflawomega}) \citep{kennicutt89,martin01,bigiel10}. Note that a radial truncation is absent for the molecular ($N=1$) star formation relation \citep{schruba11}. The possible physics behind surface density thresholds are extensively discussed by \citet{leroy08}, but we briefly summarize them here.

The efficiency of galactic star formation may be related to the global gravitational (in)stability of star-forming discs -- if the kinematics of a gas disc are such that it can withstand global collapse, star formation is suppressed \citep{toomre64,quirk72,fall80,kennicutt89,martin01}. There are indications that the surface density threshold for gravitational instability not only {applies globally}, but also locally for the spiral arms and the inter-arm regions of M51 \citep{kennicutt07}. The surface density threshold for gravitational instability in galaxy discs is based on the \citet{toomre64} $Q$ parameter:
\begin{equation}
\label{eq:Q}
Q_{\rm gas}=\frac{\sigma\kappa}{\pi G\Sigma} ,
\end{equation}
with $\sigma$ the one-dimensional gas velocity dispersion and $\kappa$ the epicyclic frequency, which is a measure for the Coriolis force in a condensing gas cloud:
\begin{equation}
\label{eq:kappa}
\kappa=\sqrt{2}\frac{V}{R}\left(1+\frac{{\rm d}{\ln{V}}}{{\rm d}{\ln{R}}}\right)^{1/2},
\end{equation}
with $V$ the circular velocity, $R$ the galactocentric radius, and $\Omega\equiv V/R$ the angular velocity. Note that for a flat rotation curve (as appropriate for the CMZ on the scales under consideration here, see \citealt{launhardt02}) we have $\kappa=\sqrt{2}\Omega$, whereas for solid-body rotation $\kappa=2\Omega$. In gas discs with $Q<1$, the self-gravity of contracting clouds is sufficient to overcome the Coriolis force and undergo gravitational collapse, whereas $Q>1$ indicates stability by kinetic support. Galaxy discs typically self-regulate to $Q\sim1$ \citep[e.g.][]{martin01,hopkins12}, with an observed range of 0.5 to~6 for galaxies as a whole \citep[e.g.][]{kennicutt89,martin01}, and an even larger variation within galaxies (see below and \citealt{martin01}). If a substantial mass fraction of the disc is constituted by stars, the disc can be unstable even though $Q_{\rm gas}>1$. In that case we write \citep[cf.][]{wang94,martin01}:
\begin{equation}
\label{eq:Qtot}
Q_{\rm tot}=Q\left(1+\frac{\Sigma_\star}{\Sigma}\frac{\sigma}{\sigma_\star}\right)^{-1}\equiv Q\psi^{-1} ,
\end{equation}
with $\sigma_\star$ the stellar velocity dispersion. More accurate definitions of $Q$ in a two-component disc exist, but the error in the above expression does not exceed 0.2~dex as long as $\sigma/\sigma_\star>0.1$ \citep{romeo11}, which holds in the CMZ (see \S\ref{sec:obs}).

If star formation is driven by global disc instability, then the inversion of equations~(\ref{eq:Q}) and~(\ref{eq:Qtot}) leads to a critical surface density for star formation, modulo a proportionality constant $\alpha_Q$:
\begin{equation}
\label{eq:surfQ}
\Sigma_{\rm Toomre}=\alpha_Q\frac{\sigma\kappa}{\pi G\psi}\equiv\frac{\sigma\kappa}{\pi GQ_{\rm crit}\psi} ,
\end{equation}
where $\psi=1$ if the contribution of stars to the gravitational potential is neglected (in the solar neighbourhood $\psi\sim1.4$, see \citealt{martin01}). \citet{kennicutt89} and \citet{martin01} empirically determined for nearby disc galaxies that $\alpha_Q=0.5$--$0.85$ with a best value of $\alpha_Q\sim0.7$, indicating that the critical Toomre $Q$ parameter for star formation is $Q_{\rm crit}=1.2$--$2$ or $Q_{\rm crit}\sim1.4$. The threshold density depends on $\sigma$ and $\kappa$, both of which increase towards the Galactic centre \citep[e.g.][]{morris96,oka01,longmore13,shetty12}, and hence it is possible that gravitational stability inhibits star formation in the CMZ.

A different, but closely related form of global disc stability to self-gravity is due to rotational shear, which may compete with self-gravity and prevent the collapse of the disc to form stars \citep{goldreich65,elmegreen87,elmegreen91,hunter98}. Rather than the epicyclic frequency of equation~(\ref{eq:surfQ}), this threshold depends on the shear time, i.e.~the time available for gas instabilities to arise during the shear-driven density growth of spirals \citep{elmegreen87,elmegreen93,elmegreen97b}. This shear condition may be more relevant than the Toomre condition if the angular momentum of a growing gas perturbation is not conserved, as might be the case in the presence of magnetic fields or viscosity. The shear time-scale is the inverse of the \citet{oort27} constant $A$:
\begin{equation}
\label{eq:oort}
A_{\rm Oort}=-0.5R\frac{{\rm d}\Omega}{{\rm d} R} ,
\end{equation}
which for a flat rotation curve becomes $A_{\rm Oort}=\Omega/2=\kappa/\sqrt{2}$ and for solid-body rotation gives $A_{\rm Oort}=0$.

The critical surface density for self-gravity to overcome shear is then
\begin{equation}
\label{eq:surfshear}
\Sigma_{\rm Oort}=\alpha_A\frac{\sigma A_{\rm Oort}}{\pi G\psi} ,
\end{equation}
with $\alpha_A\approx2.5$ \citep{hunter98}. For any rotation curve flatter than solid-body rotation (i.e.~$V\propto R^\alpha$ with $\alpha<1$), we see that $R{\rm d}\Omega/{\rm d} R$ (and hence $A_{\rm Oort}$) decreases with increasing galactocentric radius, and hence shear motion may be responsible for the suppression of star formation in the CMZ.

Before continuing, we should caution that the Toomre or shear stability of a gas disc can only act as a `soft' threshold. There are several ways in which such a threshold may be violated. For instance, small-scale structure (e.g.~spirals or bars), turbulent dissipation, magnetic stripping of angular momentum, or a soft equation of state could all lead to local star formation in a disc that is globally stable to star formation. The discussed thresholds thus represent a soft separation between star-forming and quiescent discs.

\begin{figure}
\center\resizebox{7.6cm}{!}{\includegraphics{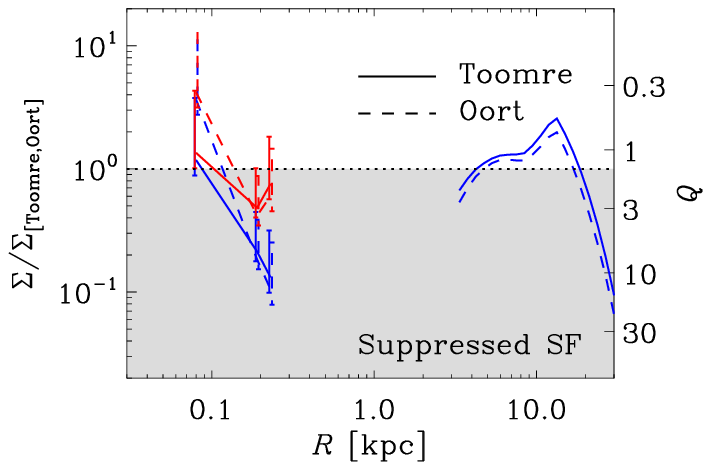}}\\
\caption[]{\label{fig:stability}
      Ratio of the gas surface density in the Milky Way $\Sigma$ to the critical surface density for star formation $\Sigma_{\{\rm Toomre,Oort\}}$ as a function of galactocentric radius. The solid and dashed lines refer to the critical densities of equations~(\ref{eq:surfQ}) and~(\ref{eq:surfshear}), respectively. The red lines include the stellar gravitational potential, whereas the blue lines exclude the contribution from stars. The range of Toomre $Q$ implied by the ratio $\Sigma/\Sigma_{\rm Toomre}$ is indicated on the right-hand side, and the error bars indicate the range of uncertainty from Table~\ref{tab:prop}. Star formation in the Galactic disc is suppressed in the grey-shaded area, i.e.~for $0.1\la R/{\rm kpc}\la4$ and $R/{\rm kpc}\ga 17$.
                 }
\end{figure}

The overdensities $\Sigma/\Sigma_{\{\rm Toomre,Oort\}}$ with respect to the surface density thresholds for gravitational instability and overcoming rotational shear are shown in Figure~\ref{fig:stability} as a function of galactocentric radius for a simple model of the Milky Way. For $R\leq0.3$~kpc, we only include the three data points from Table~\ref{tab:prop}, at galactocentric radii of $R=\{0.08,0.19,0.23\}$~kpc for the 100-pc ring, 1.3$^\circ$ complex, and the 230-pc integrated CMZ, respectively. The resulting overdensities are shown with and without the stellar contribution to the total gravitational potential. For $R\geq3$~kpc, we use the \citet{wolfire03} model for the ISM surface density profile, including a factor of 1.4 to account for the presence of helium, and adopt $\sigma=7~{\rm km}~{\rm s}^{-1}$ \citep{heiles03}. At these galactocentric radii, the epicyclic frequency $\kappa$ and the Oort constant $A$ are calculated using the rotation curve from \citet{johnston95}. Our conclusions are unaffected when using other physically appropriate Milky Way models. Finally, we correct for the presence of spiral arms by dividing $\Sigma_{\{\rm Toomre,Oort\}}$ by a factor of two \citep[c.f.][]{balbus88,krumholz05}.

Figure~\ref{fig:stability} shows that the Milky way disc is unstable to star formation in a ring covering $4\la R/{\rm kpc}\la 17$. Especially the outer edge of the star-forming disc catches the eye. The difference between the Toomre and Oort thresholds across the disc is generally too small to indicate with certainty which mechanism dominates. Only in the 100-pc ring we tentatively find $\Sigma_{\rm Toomre}>\Sigma_{\rm Oort}$, which indicates a decreasing importance of shear as the circular velocity decreases for $R<0.1$~kpc, but the difference is less than $1\sigma$. A more substantial change is brought about by including the presence of stars in the calculation of $Q$ (red lines in Figure~\ref{fig:stability}). When the stellar gravity is ignored, the CMZ outside of the 100-pc ring is highly stable to gravitational collapse with $\Sigma/\Sigma_{\{\rm Toomre,Oort\}}\sim0.1$, suggesting that star formation could potentially be suppressed. However, when the stars are included, the CMZ appears marginally stable -- the range of $Q_{\rm tot}$ measured in the CMZ is in fact very similar to that observed in normal disc galaxies \citep[compare Table~\ref{tab:prop} to][]{martin01}, which suggests a similar degree of self-regulation. The paucity of star formation in the CMZ is therefore not due to the `morphological quenching' \citep{martig09} of star formation that is considered to enable the long-term presence of quiescent gas reservoirs in galactic spheroids if $Q_{\rm tot}>1$.

The high value of $Q_{\rm gas}$ may slow down the condensation of self-gravitating gas clouds and their decoupling from the stellar background potential. The time-scale for clouds to become gravitationally unstable is $t_{\rm grav}\sim Q_{\rm gas}/\kappa$ \citep[e.g.][]{jogee05}. If the SFR is limited by the slow condensation of clouds, this therefore implies a decrease of the SFR by $1/Q_{\rm gas}$. This simple modification of the Silk-Elmegreen relation is consistent with the observed SFR for the 230~pc-integrated CMZ, but it does not explain the other components of the CMZ. Note in particular that the 100-pc ring is always marginally Toomre-stable, both in terms of $Q_{\rm gas}$ and $Q_{\rm tot}$, which is also consistent with its clumpy, beads-on-a-string morphology \citep{longmore13b}. This is likely because the ring represents a different evolutionary stage of the gas in the CMZ \citep[e.g.][]{molinari11}, which is a point we return to in \S\ref{sec:total}.

\subsubsection{Episodic star formation} \label{sec:episodic}
While it is tempting to assume that the CMZ is a steady-state system, the orbital and free-fall time-scales in the CMZ are so short (i.e.~a few Myr) that the current reservoir of dense gas may not be related to the observed star formation tracers, which originate from gas that was present at least one dynamical time-scale ago. If the star formation in the CMZ is episodic \citep[e.g.][]{loose82}, it needs to be established which physics could be driving the variability. There is a wide range of physical processes that could lead to some degree of episodicity. In this section, we discuss an extensive (but not exhaustive) selection.

\paragraph{Stochasticity} Could the low SFR simply arise from a stochastic fluctuation? At $0.015~\msun~{\rm yr}^{-1}$ \citep{longmore13}, the 100-pc ring produces about $10^5~\msun$ per dynamical time $t_{\rm dyn}\equiv2\pi/\Omega\sim4~{\rm Myr}$. {Assuming that 50~per~cent of the star formation occurs in bound clusters (as is appropriate for the CMZ, cf.~\citealt{kruijssen12d}) with a power law mass function with index $-2$ between $10^2~\msun$ and $10^6~\msun$}, this corresponds to the production of 1--2 young massive clusters (YMCs; $M\geq10^4~\msun$) and is thus consistent with the presence of the Arches and Quintuplet clusters.\footnote{We exclude the nuclear cluster of the Milky Way, of which the young stellar component is thought to have a very different origin (see e.g.~\citealt{genzel10b} and \citealt{antonini13} for recent discussions).} If the SFR were consistently an order of magnitude higher (as predicted by {density-dependent} star formation relations), then $\sim25$ such YMCs would be expected. This would imply that the present cluster population in the CMZ is a $\ga4.5\sigma$ deviation. It is thus highly unlikely that the observed SFR is due to simple Poisson noise. 

\paragraph{Statistical sampling} The CMZ agrees with the Schmidt-Kennicutt relation when averaged over a size scale of $R\sim230$~pc, which corresponds to an orbital period of 14~Myr (see Table~\ref{tab:prop}). It may be possible that above these size and time-scales, the time sequence of star formation becomes sufficiently well-sampled to correlate the present dense gas and star formation tracers \citep{kruijssen14}.\footnote{{In this context, it is interesting to note that the time-integral of the current SFRs in the 100-pc ring and the 230-pc integrated CMZ over a Hubble time does give a total stellar mass that is roughly consistent with the total stellar mass enclosed at these radii \citep[cf.][]{launhardt02}. Obviously, the bulge has not been in place for a Hubble time, but this comparison does put the present SFR in an appropriate perspective.}} We have tested this idea in \citet{kruijssen14}, finding that for the conditions of the CMZ, global star formation relations are expected to break down below 80~pc. Because this is smaller than the regions under consideration (such as the 100-pc ring), the inadequate statistical sampling of star-forming regions on small spatial scales does not explain the low SFR in the CMZ, {\it unless} there exists some physical process that synchronises the evolutionary stages of the star-forming regions in the CMZ. Possible synchronisation mechanisms could be a nuclear starburst or black hole feedback. In this scenario, the star-forming regions in the CMZ would not be independent and the CMZ could represent a single independent star-forming region, inevitably leading to phases during which its SFR exceeds or falls short of the SFR expected from global star formation relations.

\paragraph{Gravitational instability and spiral waves} Episodic star formation due to the global structural evolution of galaxy centres should be a common process if a bar is present -- in the rings of the inner Lindblad resonance, gravitational instabilities can drive fragmentation and eventually induce a starburst, but this only takes place above a certain volume density threshold \citep{elmegreen94}. If the system is steady-state on a global scale, this threshold density is given by $n_{\rm burst}=0.6\kappa^2/G\mu m_{\rm H}$ and hence $n_{\rm burst}=\{0.3,0.5,2.6\}\times10^4~{\rm cm}^{-3}$ for the three regions of the CMZ listed in Table~\ref{tab:prop}. This is comparable to the current mean density of $n_0=2\times10^4~{\rm cm}^{-3}$, which may indicate that (part of) the CMZ is currently evolving towards a starburst \citep[also see][]{oka11}.\footnote{We note that this critical density increases if the gas accretion rate along the bar is substantial.}

Additionally, orbital curvature causes the density waves in the central regions of galaxies to grow at an increasing rate towards smaller galactocentric radii \citep{montenegro99}. This does not require the gas to be Toomre-unstable or even self-gravitating, and thereby strongly contrasts with the spiral density waves in galaxy discs, which grow by gravitational instability. It is therefore easy to picture a system in which fresh gas is transported along the bar into the CMZ, where it forms spiral waves and rings, gradually building up until the critical surface density for gravitational instability and fragmentation is reached -- possibly at different times throughout the ring. There is a notable population of 24$\mu{\rm m}$ sources at $l\la359.5^\circ$, i.e.~beyond the position of Sgr~C \citep[see Figure~\ref{fig:img} and][]{yusefzadeh09}, which may be the remnant of a recent, localized starburst. Variations would typically occur on the dynamical timescale of the system, which is $\sim5$~Myr for the 100-pc ring. The runaway generation of spiral waves and the subsequent evolution towards self-gravity and instability is a promising scenario, which we will return to in \S\ref{sec:total}.

\paragraph{Feedback} The feedback of energy and momentum from young star-forming regions might also induce a fluctuating SFR. As will be discussed in \S\ref{sec:shocks}, the majority of the current star formation in the 100-pc ring takes place in the Sgr~B2 complex. In the past, the birth environment of the $24\mu{\rm m}$ sources (possibly also near the present location of Sgr~B2) may have been the dominant site of star formation. Such an asymmetry implies that the feedback from star formation originates from discrete locations. As a result, the feedback energy will escape through the path of least resistance and hence star-forming regions on one side of the ring cannot support gas on the far side (also see \S\ref{sec:rad}). This makes feedback from star formation an unlikely star formation suppressor on the spatial scales of the entire CMZ.

While a local starburst may not impact the entire 100-pc ring, it could blow out the gas in its vicinity. By the time such a burst is $\sim10~{\rm Myr}$ old, its H{\sc ii} regions will have faded and much of the gas could be in atomic form, being dispersed to higher latitudes by the combined acceleration of H{\sc ii} regions, winds, and supernovae (cf.~the giant bubbles observed in radio and $\gamma$-rays \citealt{sofue84,su10,carretti13}). Hence, the gas would be unavailable for star formation.\footnote{The gas could affect the SFR by falling back on to the CMZ, possibly supplying turbulent energy or triggering collapse. The importance of this mechanism depends on the energy injected originally by the feedback itself and is discussed in \S\ref{sec:ejecta}.} The timescale for the SFR fluctuations would then be set by the half-time of the star formation feedback. Their bolometric luminosity decreases by a factor of two in roughly 8~Myr, implying that feedback could contribute to the low-SFR `wake' after a CMZ-wide starburst, with a natural timescale of $\sim10$~Myr.

\paragraph{Gas inflow variability} Finally then, there might be a considerable time-variation of the gas inflow from large galactocentric radii onto the CMZ. Episodes of little gas inflow could then lead to depressions in the SFR relative to the available gas, provided that there is a time-delay between the presence of the gas and the onset of star formation.

\citet{kim12c} present numerical simulations of the gas flow in the central regions of barred galaxies, and measure the gas flux through a sphere with radius $R=40$~pc, i.e.~within their simulated equivalents of the 100-pc ring. They find that the variation of the gas flux is typically less than an order of magnitude in models with pronounced rings, because the inflowing gas is trapped in the nuclear ring before gradually falling to the centre. For those models that do not develop such features, and hence have an uninhibited gas flow to the sphere where the flux is measured, the fluctuations sometimes reach two orders of magnitude. This can be taken as a rough indication of the possible variation of the gas flow onto the 100-pc ring of the Milky Way, and is similar in magnitude to the present underproduction of stars in the CMZ.

The time-scale for variations of the gas flow corresponds to the dynamical timescale at the end of the bar, i.e.~$t_{\rm dyn}\sim100$~Myr, implying a relatively wide window during which the system can be observed at a low SFR. This seems at odds with current observations of the CMZ. The presence of two YMCs and the 24$\mu{\rm m}$ sources suggests substantial recent star formation activity, and considering the high densities of the several Brick-like clouds in the CMZ the current dearth of star formation is unlikely to continue.

We add one final consideration. The gas inflow along the Galactic bar may contribute to the driving of substantial turbulent motion in the CMZ (see \S\ref{sec:inflow}), which in turn may play an important role in setting a critical volume density threshold for star formation, and hence the SFR (see \S\ref{sec:turb}). Variations of the bar inflow rate may therefore indirectly affect the SFR by inducing variations in the kinetic energy budget of the gas in the CMZ.

\paragraph{Implication: a limit on YMC lifetimes} \label{sec:limit} If star formation in the CMZ is episodic, the low present SFR necessarily represents a near-minimum. The marginal Toomre stability of the ring suggest that it should collapse to form stars over the next few free-fall times and the two observed YMCs cannot provide the pressure support necessary to consistently counteract collapse throughout the CMZ. During such a burst of star formation, the present gas reservoir could produce $\sim20$ YMCs (which is the number of dense clumps observed by \citealt{molinari11}) of Arches-like masses, assuming that the progenitor clumps are similar to the cloud G0.253+0.016 \citep[also known as `The Brick', see][with $M\sim10^5~\msun$, $R\sim3~{\rm pc}$]{longmore12} and form stars in gravitationally bound\footnote{Using the cluster formation model of \citet{kruijssen12d} we find that at the high gas surface density of the 100-pc ring about $\sim50$~per~cent of the stars are expected to form in bound stellar clusters. This idea is supported by the existence of the dispersed population of $24\mu{\rm m}$ sources (see Figure~\ref{fig:img}), and the roughly equal numbers of Wolf-Rayet stars and OB supergiants observed in clusters and in the field of the CMZ \citep{mauerhan10}.} clusters at a 10--30 per cent efficiency.

Thus far, we have discussed two main physical mechanisms for episodic star formation that can act globally and locally, i.e.~the build-up of gas to a threshold for star formation and the periodic ejection of gas by feedback, respectively. Both mechanisms would act on time-scales of 5--10~Myr. This places a strong limit on the observability of any produced YMCs -- if their apparent present-day absence is not caused by their possible fading below the detection limit, their disruption timescale would have to be at most 10~Myr.

The disruption of a $10^4~\msun$ cluster in a gas-rich, high-density environment is dominated by tidal shocks due to GMC passages \citep{elmegreen10b,kruijssen11,kruijssen12c}. The dynamical friction timescale of a $10^4~\msun$ cluster orbiting in the 100-pc ring is $t_{\rm df}\sim3$~Gyr,\footnote{Using equation~(30) of \citet{antonini13}, with $\gamma=1.8$, $r_0=r_{\rm in}=80~{\rm pc}$, $\rho_0=145~\msun~{\rm pc}^{-3}$, and $m_{\rm cl}=10^4~\msun$.} implying that (on the time-scale under consideration) the YMC orbits are unaffected by dynamical friction. The YMCs will therefore keep interacting with the ring regularly after their formation.

The time-scale for disruption due to tidal shocking by GMCs is $t_{\rm dis}^{\rm sh}\propto\Sigma_{\rm GMC}^{-1}\rho_{\rm mol}^{-1}\rho_{\rm YMC}$ \citep{gieles06}, with $\Sigma_{\rm GMC}$ the GMC surface density, $\rho_{\rm mol}$ the spatially averaged molecular gas volume density, and $\rho_{\rm YMC}$ the YMC volume density. Taking the Brick as a template GMC, we have $\Sigma_{\rm GMC}\sim5\times10^3~\msun~{\rm pc}^{-2}$, whereas $\rho_{\rm mol}=\Sigma_{\rm mol}/2h=150~\msun~{\rm pc}^{-3}$ for the 100-pc ring (cf.~Table~\ref{tab:prop}). Substituting these numbers gives $t_{\rm dis}^{\rm sh}=5.7$~Myr for a $10^4~\msun$ cluster with a half-mass radius of $0.5~{\rm pc}$ \citep[cf.][]{portegieszwart10}.

The above disruption time-scale of $t_{\rm dis}^{\rm sh}=5.7$~Myr is much shorter than that found by \citet{portegieszwart01} for YMC disruption due to the smooth component of the Galactic tidal field $t_{\rm dis}^{\rm tidal}\sim80$~Myr. Based on an 80-Myr lifetime, \citet{portegieszwart01} predict the presence of 50 Arches-like clusters in the CMZ. When including tidal shocking by the dense gas, only three or four Arches-like clusters are expected to exist in the CMZ at a given time. Within the low-number Poisson statistics, this is consistent with the presence of the Arches and Quintuplet clusters. The disruption of YMCs by tidal shocks is thus capable of `hiding' some of the evidence left by previous starbursts. At 2 and 4~Myr, the ages of the Arches and Quintuplet clusters are consistent with this disruption process. {Therefore, it remains possible that star formation in the CMZ is episodic due to (1) the gradual build-up of dense gas by spiral instabilities, (2) its rapid consumption once the instability threshold is reached, and (3) a possible contribution from feedback during the post-starburst phase.}

\subsubsection{The geometry of the CMZ and tidal shocks} \label{sec:shocks}
In highly dynamical environments like the CMZ or barred galaxies, it is conceivable that GMCs are disrupted by transient tidal perturbations \citep[`tidal shocks', see e.g.][]{spitzer58,kundic95} before having been able to form stars. This idea was first introduced by \citet{tubbs82}, who suggested that perturbations would limit the duration of star formation and hence decrease the SFR.\footnote{Note that despite the similarity in nomenclature, a tidal shock bears no physical resemblance to a hydrodynamical shock. Cloud destruction by tidal shocking also differs from cloud disruption by a steady tidal field (see e.g. \citealt{kenney92} for extragalactic examples), to which we turn in \S\ref{sec:tidal}.}

The vast majority of the star formation in the 100-pc ring of the CMZ takes place in the complex Sgr~B2 \citep{longmore13}. It has been proposed by \citet{molinari11} that the 100-pc ring coincides with the $x_2$ orbit -- a family of elliptical orbits with semi-major axis perpendicular to the Galactic bar, which precesses at the same rate as the pattern speed of the bar, resulting in stable, non-intersecting trajectories. The $x_2$ orbits are situated within the $x_1$ orbits, which are elongated along the major axis of the bar \citep[see e.g.][]{athanassoula92}. In this scenario, Sgr~B2 and Sgr~C lie at the points where the $x_1$ and $x_2$ orbits touch. The superposition of both orbital families then leads to the accumulation of gas, generating two standing density waves at the locations of Sgr~B2 and Sgr~C with pattern speeds different from the flow velocity of the gas orbiting on the 100-pc ring. Such a configuration has been observed in other galaxies \citep[e.g.][]{pan13}.

\begin{figure}
\center\resizebox{\hsize}{!}{\includegraphics{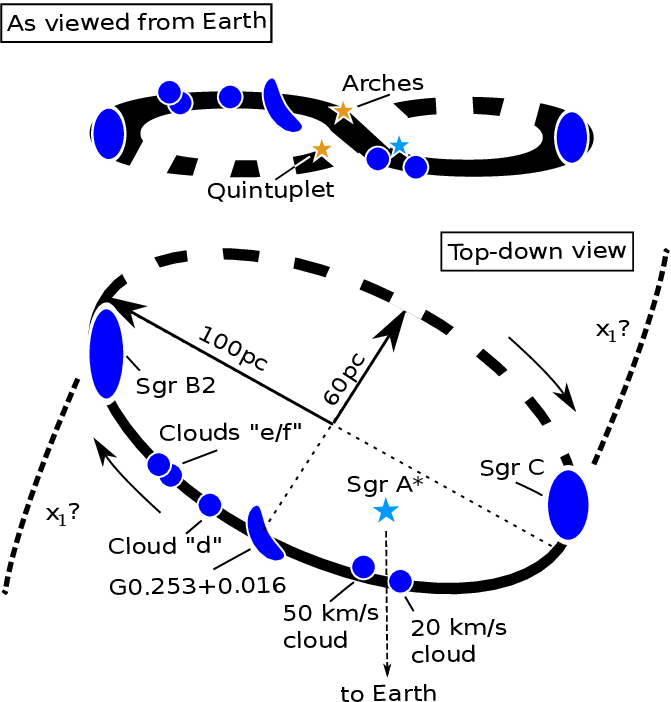}}\\
\caption[]{\label{fig:cmzschem}
      {Schematic representation of the picture discussed in \S\ref{sec:shocks} \citep[also see][]{bally10,molinari11,longmore13b}, as seen from Earth (top) and from above (bottom). Note that the position along the line of sight of the Arches and Quintuplet clusters is unknown, and therefore they are not included in the top-down view. As discussed in \S\ref{sec:globalobs} and footnote~\ref{footnote:uncertain}, it is possible that the ring extends further, in which case the $x_1$ orbits may connect to the ring under a different angle, and at a different galactocentric radius. The discussion of \S\ref{sec:shocks} assumes the geometry depicted here to put upper limits on cloud disruption (see text).}
                 }
\end{figure}

The ring upstream of Sgr~B2 is fragmented into clouds that have properties suggesting that they could be massive protoclusters \citep{longmore13b}. A schematic representation of this configuration is shown in Figure~\ref{fig:cmzschem}. If Sgr~B2 is indeed a standing density wave and the impending encounter of these clouds with Sgr~B2 is sufficiently energetic \citep[cf.][]{sato00}, it is possible that star formation is briefly induced due to the tidal compression when they enter Sgr~B2, but is subsequently halted when the clouds exit the Sgr~B2 region and rapidly expand due to having been tidally heated. We reiterate that the scenario of Figure~\ref{fig:cmzschem} is not undisputed (see footnote~\ref{footnote:uncertain}). However, it does lead to the most extreme cloud-cloud encounters that could take place in the CMZ, and therefore we consider it as an upper limit to their disruptive potential.

It is straightforward to quantify the disruptive effect of a tidal perturbation as the ratio of the energy gain $\Delta E$ to the total energy of the cloud $E$, under the condition that the duration of the perturbation is shorter than the dynamical time of the perturbed cloud \citep[the `impulse approximation', see][]{spitzer87} -- otherwise the injected energy is simply dissipated. In order to compute the relative energy gain, we consider a head-on collision and approximate the cloud and the perturber with \citet{plummer11} potentials. If we include the correction factors for the extended nature of the perturber \citep{gieles06} and the second-order energy gain \citep{kruijssen11}, the total relative energy gain becomes:
\begin{equation}
\label{eq:deshock}
\left|\frac{\Delta E}{E}\right|=1.1M_7^2 R_{{\rm h},1}^{-4} m_5^{-1} r_{{\rm h},0}^3 V_2^{-2} ,
\end{equation}
where $M_7\equiv M/10^7~\msun$ is the perturber mass, $R_{{\rm h},1}\equiv R_{\rm h}/10~{\rm pc}$ is its half-mass radius, $m_5\equiv m/10^5~\msun$ is the cloud mass, $r_{{\rm h},0}\equiv r_{\rm h}/1~{\rm pc}$ is its half-mass radius, and $V_2\equiv V/100~{\rm km}~{\rm s}^{-1}$ is the relative velocity between both objects. The cloud is unbound if $|\Delta E/E|\geq1$.

To describe Sgr~B2 we adopt $M_7=0.6$ \citep{bally88,goldsmith90} and $R_{{\rm h},1}=1.5$, for the clouds approaching Sgr~B2 we use the properties of the Brick, with $m_5=1.3$ and $r_{{\rm h},0}=2.8$, and the relative velocity is taken to be $V_2\sim1$ (slightly higher than the line-of-sight streaming velocity in the 100-pc ring). For these numbers, equation~(\ref{eq:deshock}) gives $|\Delta E/E|=1.3$, suggesting that a Brick-like cloud could in principle {\it just} be unbound when passing through the gravitational potential chosen to represent Sgr~B2. To verify the validity of the impulse approximation, we note that the duration of the perturbation is $\Delta t=2R_{\rm h}/V=0.1$--$0.2$~Myr, whereas the dynamical time is $t_{\rm dyn}=(G\rho_{\rm h})^{-1/2}\sim0.4$~Myr, and hence $\Delta t<t_{\rm dyn}$.

The above approach does not account for the detailed structure of Sgr~B2 and the passing clouds, nor does it cover the collisional hydrodynamics of the clouds \citep[e.g.][]{habe92}, their dissipative nature, or any possible, substantial deviations from spherical symmetry. Most of these effects would weaken the disruptive effect of the tidal perturbation. The collisional hydrodynamics of the system imply that, depending on the ram pressure balance of the interaction, a substantial fraction of the Brick will lag behind its ballistic orbit and may never emerge from the possible density wave at Sgr~B2. Upon entering the density wave, the cloud will be compressed and the turbulent energy dissipation rate is enhanced accordingly -- dissipating the tidally injected energy along the way. Finally, any deviation from spherical symmetry implies that the cloud collapses more rapidly than its spherically symmetric idealisation \citep{pon12}, and hence the compression when entering the perturbation may trigger runaway collapse.

We conclude that the unique geometry of the CMZ may affect star formation in part of the region, but it is unlikely that it would inhibit star formation. The hydrodynamic perturbation of the clouds passing through a density wave likely accelerates their collapse, in which case Sgr~B2 and Sgr~C represent the instigation points of star formation rather than the loci where it is terminated. Furthermore, this model may apply to star formation in the 100-pc ring, but it remains to be seen to what extent similar effects could affect star formation elsewhere in the CMZ.

\subsection{Locally inhibited star formation} \label{sec:local}
We now turn to a discussion of the possible mechanisms on spatial scales smaller than the disc scale height ($\Delta R<h$) that may limit star formation in the CMZ. Volumetric star formation relations generally rely on the free-fall time $t_{\rm ff}\propto n^{-1/2}$ to predict $\Sigma_{\rm SFR}$, because a few per~cent of the gas mass is converted into stars per dynamical time or free-fall time \citep{elmegreen02,krumholz07,evans09}. While this number and its time-evolution in individual clouds is still debated \citep{padoan14}, the relative universality of its mean value suggests that a similar fraction of the gas mass in molecular clouds ends up participating in star formation, irrespective of the environmental conditions. In star formation theories, this is explained by the idea that the dispersion of the volume density PDF depends on the Mach number in a similar way to the critical volume density for star formation -- the PDF broadens as the density threshold increases \citep[e.g.][]{krumholz05,padoan11}. Considering the broad range of gas densities observed in disc and starburst galaxies, this is a crucial ingredient to allow the SFE per free-fall time to be roughly constant. 

The above picture has been challenged by the recent observation by \citet{lada10,lada12} that there is a possibly universal, critical volume density $n_{\rm Lada}\sim10^4~{\rm cm}^{-3}$ for converting gas into stars. A threshold density for star formation should be expected, as there obviously exists some density above which all gas ends up in stars (modulo the mass lost by protostellar outflows), but it is not clear why such a transition density would be universal. As discussed in \S\ref{sec:obs}, the SFR currently observed in the CMZ implies that the gas mass fraction that is used to form stars is $f_{\rm th}\la0.001$. We showed in \S\ref{sec:obs} that if a volume density threshold for star formation exists, its value in the CMZ has to be $n_{\rm th}\geq10^7~{\rm cm}^{-3}$ for various parametrizations of the density PDF, i.e.~$n_{\rm th}=10^7$--$3\times10^8~{\rm cm}^{-3}$ for a power-law approximation and $n_{\rm th}\sim\{0.3,1\}\times10^9~{\rm cm}^{-3}$ for a log-normal when including and excluding the effect of the magnetic field, respectively.

In the following, we verify which physical mechanisms are consistent with the inhibition of star formation below such densities. Because the turbulent pressure in the CMZ is remarkably high \citep[e.g.][]{bally88}, with $P_{\rm turb}/k=\mu m_{\rm H}n\sigma^2/k\sim10^9~{\rm K}~{\rm cm}^{-3}$, an important constraint is that potential star formation inhibitors should be able to compete with the turbulence. Therefore, we often use the turbulent pressure as a reference point to calculate the gas volume densities below which the SFR may be suppressed by each mechanism. The mechanism responsible for the low observed SFR needs to be effective up to a critical density of $n_{\rm th}\geq10^7~{\rm cm}^{-3}$.

\subsubsection{Galactic tides} \label{sec:tidal}
A first condition for initiating star formation is that the progenitor clouds are not tidally disrupted. The gas needs to have a volume density higher than the tidal density, i.e.~the density required for a spherical density enhancement to remain bound in a galactic tidal field. While this does not guarantee that a cloud will eventually form stars, it does represent a key requirement for star formation to proceed. The tidal density is written as
\begin{equation}
\label{eq:ntidal}
n_{\rm tidal}=\frac{3A_{\rm pot}\Omega^2}{4\pi \mu m_{\rm H}G} ,
\end{equation}
which only depends on the angular velocity $\Omega$. This expression assumes that the cloud orbit is circular. The constant $A_{\rm pot}$ depends on the shape of the galactic gravitational potential, and is $A_{\rm pot}=\{0,2,3\}$ for solid-body rotation, a flat rotation curve, and a point source (i.e.~Keplerian) potential, respectively. {Adopting the \citet{launhardt02} potential for the CMZ, we find $A_{\rm pot}=\{2.0,1.9,0.7\}$ for the 230~pc-integrated CMZ, the 1.3$^\circ$ complex, and the 100-pc ring, respectively.} Galactic tides inhibit star formation in regions where $n_{\rm tidal}>n_{\rm th}$, with $n_{\rm th}$ some unknown threshold for star formation. For the three regions of the CMZ that are listed in Table~\ref{tab:prop}, equation~(\ref{eq:ntidal}) yields $n_{\rm tidal}=\{1.3,1.8,2.1\}\times10^3~{\rm cm}^{-3}$. This is several orders of magnitude lower than the $n_{\rm th}\geq10^7~{\rm cm}^{-3}$ required to explain the SFR in the CMZ. It is also only an order of magnitude lower than the mean gas volume density in the CMZ, indicating that it may affect the accretion of gas onto molecular clouds, but not the star formation within them.

We have not accounted for eccentric orbits, because even if tides only affected the cloud structure (i.e.~$n_{\rm tidal}\geq n_{\rm cloud}$) rather than the star formation process itself (i.e.~$n_{\rm tidal}\geq n_{\rm th}$), this would require such a high angular velocity ($\Omega_{\rm crit}\ga10~{\rm Myr}^{-1}$) that even at fixed circular velocity an eccentricity of $\epsilon\geq0.9$ would be necessary for tides to be the limiting factor. Such extreme eccentricities are ruled out by recent dynamical models of the orbital structure of the gas in the CMZ (Kruijssen, Dale \& Longmore, in~prep.). We conclude that star formation in the CMZ is not inhibited by tides.

\subsubsection{Turbulence} \label{sec:turb}
Turbulence plays a key role in the recent star formation models of \citet{krumholz05} and \citet{padoan11}. While the former proposes that turbulent pressure support sets the critical volume density for star formation on the sonic scale, the latter work takes the point that turbulence is only responsible for driving local gravitational instabilities, and that the critical volume density for star formation is set by the thickness of the post-shock layers in the supersonic ISM. These differences aside, both models do predict a critical overdensity $x\equiv n/n_0$ for star formation\footnote{For the \citet{krumholz05} model this assumes a typical size-linewidth relation for GMCs of $\sigma\propto R^{0.5}$.} that scales with the GMC virial parameter $\alpha_{\rm vir}$ and Mach number as
\begin{equation}
\label{eq:xturb}
x_{\rm turb}=A_x\alpha_{\rm vir}{\cal M}^2 .
\end{equation}
Both models also have remarkably similar proportionality constants $A_x\sim1$ to within a factor of 1.5.

Adopting a virial parameter of $\alpha_{\rm vir}=1.5$ \citep[cf.][]{krumholz05,padoan11},\footnote{Note that the clouds in the CMZ outside of the 100-pc ring have $\alpha_{\rm vir}>1.5$ when the stellar gravitational potential is omitted.} a Mach number of ${\cal M}=30$ as in \S\ref{sec:obs}, and a mean density of $n_0=2\times10^4~{\rm cm}^{-3}$, we see that the critical density for star formation is $n_{\rm turb}\equiv x_{\rm turb}n_0\sim3\times10^7~{\rm cm}^{-3}$. This approaches the low end of the required threshold densities $n_{\rm th}\sim\{1,4\}\ga10^7~{\rm cm}^{-3}$ (see Figure~\ref{fig:nth}). If we apply equation~(\ref{eq:xturb}) to the Brick \citep{longmore12}, with $\alpha_{\rm vir}\sim1$, ${\cal M}\sim20$ \citep{rathborne14}, and $n_0=7.3\times10^4~{\rm cm}^{-3}$, we obtain $n_{\rm turb}\sim4\times10^7~{\rm cm}^{-3}$. By contrast, the typical properties of GMCs in the solar neighbourhood are $\alpha_{\rm vir}\sim1.5$, ${\cal M}\sim10$, and $n_0=10^2~{\rm cm}^{-3}$, which gives $n_{\rm turb}\sim1.5\times10^4~{\rm cm}^{-3}$. Interestingly, this equals the \citet{lada10} threshold for star formation to within the uncertainties of this calculation.

The above numbers are very suggestive, in that these physically-motivated overdensity thresholds for star formation correctly predict the Lada threshold for star formation in the solar neighbourhood, as well as a density threshold in the CMZ of $n_{\rm turb}\ga3\times10^7~{\rm cm}^{-3}$, which gives the best agreement with the observed SFR so far. This threshold is a lower limit, because the CMZ clouds outside the 100-pc ring are supervirial when the stellar gravitational potential is omitted and hence have $\alpha_{\rm vir}>1.5$ (cf.~equation~\ref{eq:xturb}). This is easily offset by uncertainties in the density PDF, in the above numbers, and/or in the observed SFR.

While turbulence is capable of increasing the density threshold to the required, extreme densities, we should note that this argument is incomplete, as we have not established what is driving the turbulence. We return to this point in \S\ref{sec:disc}, and will now briefly discuss a particularly interesting uncertainty.

\subsubsection{A bottom-heavy initial mass function} \label{sec:imf}
Observational measures of the SFR are strongly biased to the emission from massive stars ($m\ga8~\msun$). The low observed SFR in the CMZ may therefore be spurious {if there is an overproduction of low-mass stars with respect to the observed massive stars. This would affect all methods of measuring the SFR, whether} it is determined using massive YSOs \citep{yusefzadeh09} or the ionising flux from massive stars \citep{murray10b,lee12}. Neither technique is capable of reliably detecting stars below $8~\msun$. Recent observational studies of giant elliptical galaxies have found evidence for a bottom-heavy IMF ${\rm d} n/{\rm d} m\propto m^{-\beta}$ with a power-law slope of $\beta=3$ at the low-mass end \citep[$m\la1~\msun$, see e.g.][]{vandokkum10,cappellari12,goudfrooij13}.

It has been suggested that the characteristic mass scale of the core mass function (CMF) is set by the thermal Jeans mass $m_{\rm J}\propto T^{3/2}n^{-1/2}$ \citep{elmegreen08b} or the sonic mass $m_{\rm sonic}$ \citep{hopkins12b}. The Jeans mass is insensitive to the volume density and the radiation field if the Schmidt-Kennicutt relation is satisfied \citep{elmegreen08b}. {However, the SFR density in the CMZ falls below the Schmidt-Kennicutt relation -- the corresponding lack of heating should lead to a lower thermal-to-total energy density ratio than in the Galactic disc (in other words, the Mach number is higher than in the disc)} and consequently the thermal Jeans mass is low too. This makes the CMZ a prime example of an environment where the CMF could have a lower-than-normal peak mass.

If we assume that the CMF and IMF are related, the above line of reasoning would imply that the characteristic turnover mass of the IMF is environmentally dependent. In the solar neighbourhood it is observed to be $m\sim0.5~\msun$ \citep{kroupa01,chabrier03}. Because $m_{\rm J}$ and $m_{\rm sonic}$ decrease with the pressure, this would result in an enhanced population of low-mass stars ($m<0.5~\msun$) in the vigorously star-forming galaxies observed at high redshift \citep[e.g.][]{vandokkum04,daddi07}, which reach Mach numbers of ${\cal M}\sim100$ \citep[e.g.][]{swinbank11} and may be the progenitors of current giant elliptical galaxies. Such Mach numbers strongly contrast with the ${\cal M}\sim10$ in the Milky Way disc. 

For core masses $M\la1~\msun$ (i.e.~stellar masses $M\la0.5~\msun$) and a Mach number of ${\cal M}\sim100$, \citet{hopkins12b} predicts a mass spectrum with a slope of $\beta\sim3$, which is much steeper than observed in the solar neighbourhood for the same mass range \citep[$\beta\sim1.5$, e.g.][]{chabrier03}. Using parameters that are appropriate for the Brick in the CMZ \citep{longmore12} and adopting a SFE in protostellar cores of $\epsilon=0.5$ \citep{matzner00}, we find $\epsilon m_{\rm sonic}\sim0.01~\msun$, whereas in the solar neighbourhood $\epsilon m_{\rm sonic}\sim0.5~\msun$. Similarly, at the approximate density threshold for star formation due to turbulence $n_{\rm turb}\sim10^9~{\rm cm}^{-3}$, the thermal Jeans mass in the CMZ ($T=65$~K) is about $\epsilon m_{\rm J}\sim0.1~\msun$, whereas in the solar neighbourhood it is $\epsilon m_{\rm J}\sim0.5~\msun$. These low characteristic masses suggest that the CMZ might be the low-redshift equivalent to the progenitor environment of giant elliptical galaxies \citep{kruijssen13}, and it is therefore important to verify to what extent any unseen stellar mass at $m<0.5~\msun$ may increase the SFR inferred for a `normal' IMF \citep{kroupa01,chabrier03}.

We integrate the mass of a \citet{kroupa01} IMF between $m_{\rm min}=0.08~\msun$ and $m_{\rm max}=100~\msun$ and compare it to the mass integral of a similar IMF, but with a power-law slope of $\beta=3$ for masses $m<0.5~\msun$. This increases the total mass by a factor of two, at the same number of massive stars. We conclude that while this is a non-negligible factor, it is (1) comparable to the uncertainty on the SFR in the CMZ and (2) in itself insufficient to explain the factor of $\geq10$ suppression of the SFR in the CMZ.

\subsubsection{The atomic-molecular phase transition of hydrogen} \label{sec:phase}
Dense molecular gas as traced by HCN is found to be correlated with star formation tracers \citep[e.g.][]{gao04,wu05}. If this relation is causal in nature, molecular gas may be required to form stars (e.g.~\citealt{schruba11}, although see \citealt{glover12} and \citealt{krumholz12c} for an alternative view). In that case, the low SFR in the CMZ may be caused by the existence of a star formation threshold due to the phase transition of H{\sc i} to H$_2$ \citep{blitz04,krumholz09c}. At solar metallicity, this transition occurs at $\Sigma\sim10~\msun~{\rm pc}^{-2}$.

The idea that the presence of molecular gas determines the SFR is a corollary of the \citet{bigiel08} relation, i.e.~${\rm SFR}\propto M_{\rm mol}$, where $M_{\rm mol}$ is the molecular hydrogen mass. This is the only relation that predicts a constant SFR per unit molecular mass and thereby fits the SFR in the CMZ (see \S\ref{sec:obs}). However, the surface density scale for the phase transition decreases with increasing metallicity \citep{krumholz09b}, and hence should be even lower than $10~\msun~{\rm pc}^{-2}$ in the central bulge region \citep[e.g.][]{brown10}. A quick look at Table~\ref{tab:prop} reveals that gas in the CMZ resides at much higher surface densities and indeed it is observed to be molecular \citep[e.g.][]{morris96}. We therefore rule out the atomic-to-molecular transition at $\Sigma\la10~\msun~{\rm pc}^{-2}$ as the cause for a suppressed SFR in the CMZ.

\subsubsection{The Galactic magnetic field} \label{sec:magnetic}
Another possible explanation would be that the high magnetic field strength in the CMZ \citep[$B\ga100\mu{\rm G}$,][]{crocker10} inhibits star formation \citep[e.g.][]{morris89}. Using the condition that the magnetic and turbulent pressure are balanced, this implies a critical volume density of
\begin{equation}
\label{eq:nmag}
n_{\rm mag}=\frac{1}{2\mu_0 \mu m_{\rm H}}\left(\frac{B}{\sigma}\right)^2 ,
\end{equation}
where $\mu_0=4\pi$ is the vacuum permeability constant. {Using the values for each of the three CMZ regions from Table~\ref{tab:prop}, this gives a critical density of $n_{\rm mag}\sim50~{\rm cm}^{-3}$ above which the magnetic pressure becomes less than the turbulent pressure.} Therefore, the magnetic field cannot provide support against the turbulence in the CMZ. This result is unchanged when adopting the internal velocity dispersions of the clouds in the CMZ rather than the large-scale velocity dispersion (in which case $n_{\rm mag}\sim100~{\rm cm}^{-3}$). A magnetic field strength of $B\ga2~{\rm mG}$ would be necessary for the magnetic pressure to compete with the global turbulent pressure, i.e.~$n_{\rm mag}\sim n_0=2\times10^4~{\rm cm}^{-3}$.

It is important to note that the measured magnetic field strength of $B\ga100\mu{\rm G}$ applies to the low-density intercloud medium, and may be an order of magnitude higher in dense clouds \citep{morris06}. This could increase the critical density to $n_{\rm mag}\sim10^4~{\rm cm}^{-3}$, which is comparable to the mean gas density, but is still much lower than the density threshold required by density-dependent star formation relations. While it does not inhibit star formation directly, the presence of a $100\mu{\rm G}$ magnetic field is likely important in shaping the properties of the ISM on the cloud scale. On the one hand, it is capable of narrowing the density PDF of the ISM in the CMZ somewhat, and slows down star formation accordingly (see \S\ref{sec:localobs} and \S\ref{sec:turb}). On the other hand, magnetic breaking leads to angular momentum loss during the condensation and contraction of cores, and hence accelerates star formation \citep{elmegreen87}.

\subsubsection{Radiation pressure} \label{sec:rad}
Considering the high angular velocity of the gas in the CMZ, the 100-pc ring needs to enclose some $10^9~\msun$ of mass. If we make the reasonable assumption that most of this mass is constituted by stars, this implies a high stellar surface density of the CMZ of $\Sigma_\star\sim3\times10^3~\msun~{\rm pc}^{-2}$ within the gas disc scale height (see Table~\ref{tab:prop}). It is therefore worth investigating whether stellar feedback is capable of inhibiting star formation in the CMZ. After the birth of a stellar population, feedback is first dominated by protostellar outflows, followed by radiative feedback, supernovae, and stellar winds. The relative importance of these mechanisms depends on the spatial scale and the environment. It has been shown by \citet{murray10} that in all but the lowest-density environments (e.g.~ GMCs in the Galactic disc) radiation pressure is the dominant feedback mechanism for disrupting GMCs,\footnote{Although see \citet{lopez13} for a counterexample.} whereas on scales $\ga100$~pc the energy deposition by supernovae becomes important. Crucially though, these mechanisms each generate a similar total energy output.

We now test the hypothesis that stellar radiation inhibits star formation in the CMZ. As discussed in \S\ref{sec:episodic}, this scenario has the problem that the CMZ hosts discrete star formation events, implying that the feedback on one side of the CMZ may not be able to affect gas on the opposite side. To test whether feedback is a viable explanation from an energy perspective, we again require pressure equilibrium between radiation pressure and turbulent pressure, which yields a critical stellar surface density of young stars $\Sigma_{\star,{\rm rad}}$ above which radiation pressure is important:
\begin{equation}
\label{eq:stelcrit}
\Sigma_{\star,{\rm rad}}=\frac{4c\mu m_{\rm H}n\sigma^2}{\Psi (1+\phi_{\rm tr}\kappa_0T^2\Sigma)} ,
\end{equation}
where $c$ is the speed of light, $\Psi\sim3\times10^3~{\rm erg}~{\rm s}^{-1}~{\rm g}^{-1}\sim1.5\times10^3~{\rm L}_\odot~\msun^{-1}$ the light-to-mass ratio of a young, well-sampled stellar population \citep{thompson05}, and the term in parentheses indicates the optical depth $1+\phi_{\rm tr}\tau$ with $\tau\sim\kappa_R\Sigma\sim\kappa_0T^2\Sigma$, in which $\phi_{\rm tr}\equiv f_{\rm tr}/\tau\sim0.2$ \citep{krumholz12b} is a constant that indicates the fraction of infrared radiation that is trapped at an optical depth $\tau=1$, $\kappa_{\rm R}$ is the Rosseland mean dust opacity \citep[cf.][]{thompson05,murray10}, and $\kappa_0\sim2.4\times10^{-4}~{\rm cm}^2~{\rm g}^{-1}~{\rm K}^{-2}$ is a proportionality constant.

We consider two cases for calculating the critical stellar surface density above which radiation pressure can compete with the turbulent pressure.  In both cases, we assume $T=65~{\rm K}$ as in \S\ref{sec:obs}, which yields an infrared optical depth of $\tau\sim\kappa_0T^2\Sigma\sim1$.

Firstly, in the 100-pc ring we have $n_0\sim2\times10^4~{\rm cm}^{-3}$, $\sigma\sim15~{\rm km}~{\rm s}^{-1}$ and $\Sigma=3.0\times10^3~\msun~{\rm pc}^{-2}$. Substituting these numbers into equation~(\ref{eq:stelcrit}) gives $\Sigma_{\star,{\rm rad}}\sim2.9\times10^4~\msun~{\rm pc}^{-2}$. Combining this with the surface area taken up by the gas in the ring ($5\times10^3~{\rm pc}^2$) and the lifetime of strongly radiating stars \citep[$\sim4~{\rm Myr}$,][]{murray10}, we see that the 100-pc ring requires a SFR of $\sim40~\msun~{\rm yr}^{-1}$ for radiation pressure to overcome the turbulence. This is over three orders of magnitude higher than the measured $\sim0.015~\msun~{\rm yr}^{-1}$.

The second case we consider is the GMC `the Brick' \citep{longmore12}, for which we adopt $n\sim7.3\times10^4~{\rm cm}^{-3}$, $\sigma\sim10~{\rm km}~{\rm s}^{-1}$ and $\Sigma=5.3\times10^3~\msun~{\rm pc}^{-2}$. The Brick requires a similar surface density of  $\Sigma_{\star,{\rm rad}}\sim4.3\times10^4~\msun~{\rm pc}^{-2}$. Other than implying a critical SFR of $0.25~\msun~{\rm yr}^{-1}$, it also means that a SFE of $\sim8$ (not per~cent!) is required to overcome the turbulent pressure, {unless the newly formed stellar population has a three times smaller radius than its parent cloud, in which case a SFE of unity would imply similar turbulent and radiative pressures.} The above numbers change by a relatively small amount when also including the combined flux of the more numerous, old stars of the Galactic bulge. Because radiative feedback is only capable of affecting the gas in the direct vicinity of highly concentrated clusters of young stars, it is unable to stop the entire CMZ from forming stars (also see~\ref{sec:episodic}).

\subsubsection{Cosmic rays} \label{sec:rays}
Cosmic ray pressure could be important in the CMZ, due to past star formation events or black hole activity. For instance, it is possible that star formation in the CMZ is inhibited by the cosmic ray flux from supernovae (SNe). We equate the cosmic ray pressure due to SNe to the turbulent pressure and assume the extreme case in which the cosmic ray energy remains trapped in the CMZ. This yields the critical volume density above which turbulence overcomes the cosmic ray pressure:
\begin{equation}
\label{eq:ncr}
n_{\rm cr}=\frac{\eta_{\rm SN}E_{\rm SN}\Gamma_{\rm SN}\tau_{\rm cr}}{\mu m_{\rm H}V\sigma^2} .
\end{equation}
In this expression, $\eta_{\rm SN}\sim0.3$ is the fraction of the SN energy that is converted to cosmic rays, $E_{\rm SN}\sim10^{51}~{\rm erg}$ is the energy of a single SN, $V$ is the volume, and $\Gamma_{\rm SN}$ is the SN rate, which for a \citet{kroupa01} or \citet{chabrier03} initial mass function (IMF) is given by $\Gamma_{\rm SN}=0.01~{\rm yr}^{-1}~({\rm SFR}/\msun~{\rm yr}^{-1})$, where the SFR is expressed in $\msun~{\rm yr}^{-1}$. The variable $\tau_{\rm cr}\propto n^{-1}$ indicates the lifetime of cosmic rays to energy loss via collisions with protons in the gas, which is $\tau_{\rm cr}\sim10^8/n_0~{\rm yr}\sim5\times10^3~{\rm yr}$.

Based on Table~\ref{tab:prop}, we have SFRs of $\{3.3,0.17,1.5\}\times10^{-2}~\msun~{\rm yr}^{-1}$ and volumes $V\sim\{17,0.89,0.050\}\times10^6~{\rm pc}^3$ for the 230~pc-integrated CMZ, the 1.3$^\circ$ complex, and the 100-pc ring, respectively. We adopt a single velocity dispersion of $\sigma\sim15~{\rm km}~{\rm s}^{-1}$ and substitute these numbers into equation~(\ref{eq:ncr}). In the three environments under consideration, this gives densities $n_{\rm cr}=\{0.1,0.1,17\}~{\rm cm}^{-3}$ above which turbulent pressure outweighs the cosmic ray pressure. Each of these densities is much lower than the mean volume density of the molecular gas in the CMZ.

Given these numbers, it is very unlikely that cosmic rays from SNe affect the gas dynamics in the CMZ. For the cosmic ray pressure to compete with turbulence dynamically, the SFR would need to be at least three orders of magnitude higher than is being observed. Even if all SN energy would be converted to cosmic rays, and their lifetime would be a hundred times longer, cosmic rays would still imply a critical volume density $n_{\rm cr}$ smaller than the mean density $n_0$. In addition, cosmic rays may be removed from the CMZ by the galactic wind before they can reach the bulk of the dense molecular hydrogen \citep{crocker11}. Our conclusion is supported by observational constraints on the cosmic ray pressure in the CMZ, which is around $P_{\rm cr}\sim10^{-10}~{\rm erg}~{\rm cm}^{-3}$ \citep{crocker11b} and hence implies $n_{\rm cr}\sim10~{\rm cm}^{-3}$, which is within a factor of a few of our theoretical derivation.

Alternatively, cosmic rays could originate from the activity of the central black hole of the Milky Way, which accretes at a rate of $\dot{M}\la10^{-8}~\msun~{\rm yr}^{-1}$ \citep{quataert00,baganoff03}. We assume that 0.5~per~cent of the accreted rest mass energy is available to heat the gas \citep[e.g.][]{dimatteo05}, and follow a similar argument as for the case of SN-powered cosmic rays above. This yields a critical volume density of $n_{\rm BH}\sim3\times10^{-2}~{\rm cm}^{-3}$ above which turbulent pressure outweighs the accretion-generated cosmic ray pressure in the volume of the 100-pc ring, again much lower than the mean volume density of the molecular gas in the CMZ.

Cosmic rays are also unimportant compared to the thermal pressure anywhere other than the 100-pc ring, where the thermal and cosmic ray pressures are comparable to within a factor of a few. Hence, they may not be important kinematically, but they could be relevant for setting the temperature of the gas \citep{ao13,yusefzadeh13}. We conclude that feedback processes in general, and radiative, supernova, and black-hole feedback in particular, are not consistently inhibiting star formation in the CMZ.\footnote{Although locally they may drive arches and bubbles -- there will always be some (small) volume $V$ such that the feedback pressure from the enclosed stars competes with (or outweighs) the turbulent pressure. For the 100-pc ring (and hence Sgr~B2 where most of the star formation takes place), cosmic rays overcome the turbulence for $V\la50~{\rm pc}^{3}$ or $R\la2~{\rm pc}$.} Of course, whether or not this also holds in other, extragalactic cases depends on their recent star formation history and their black hole activity.

\section{Implications and predictions for future observations of the CMZ} \label{sec:impl}
We now turn to the implications and possible tests of the remaining plausible star formation inhibitors of \S\ref{sec:thresholds}, which are summarized as follows. On global scales, star formation could be episodic due to the gradual build-up of dense gas by spiral instabilities and its rapid consumption once the density threshold for gravitational instability is reached. On local scales, the reduced SFR is consistent with an elevated volume density threshold for star formation due to the high turbulent pressure in the CMZ. This solution could be aided by most of the other potential star formation inhibitors that each individually were shown to be insufficient to cause the low SFR.

\subsection{Testing episodic star formation} \label{sec:testepisodic}
Evidence exists of episodic star formation events in the CMZ \citep{sofue84,blandhawthorn03,yusefzadeh09,su10} and mechanisms have been proposed to explain how such episodicity can occur. As discussed in \S\ref{sec:episodic}, instabilities can drive the fragmentation of the nuclear ring and eventually induce a starburst. Gas in barred spiral galaxies like the Milky Way is funnelled from the disc through the bar to the galaxy centre \citep{sakamoto99,kormendy04,sheth05}. Because the conditions in the CMZ lead to a higher threshold for star formation than in the disc, the gas needs time to accumulate before initiating star formation. While this by itself can already cause star formation to be episodic, it is also known from numerical simulations \citep{hopkins10b} that the presence of a bar can cause substantial variations of the gas inflow towards a galaxy centre. The large variation of the central gas concentration of otherwise similar galaxies sketches a similar picture \citep{sakamoto99}.

To constrain the possible episodicity of star formation in the CMZ, it will be necessary to map the structure of the gas flow along the Galactic bar, which is already possible using sub-mm and radio surveys of the Galactic plane \citep[e.g.][]{molinari10,walsh11,purcell12}. If the CMZ is presently near a low point of an episodic star formation cycle, then the gas needs to be accumulating and hence the inflow rate has to exceed the SFR. An improved 6D (position-velocity) map of the CMZ itself would also help to understand the nature of the possible, large-scale instabilities of the gas -- the combination of line-of-sight velocities, proper motions, plane-of-the-sky positions, and X-ray light echo timing measurements should lead to a conclusive picture of gas inflow, accumulation, and consumption.

The end result of the star formation process should also be considered further. With infrared data and spectral modelling, it is essential to establish the recent ($\sim100~{\rm Myr}$), spatially resolved star formation history of the CMZ. If evidence is found for a statistically significant variation of the SFR, the time-scale of such variations can be used to determine whether stellar feedback, gas instabilities, or a varying gas inflow rate are responsible (although see \S\ref{sec:driver}--\ref{sec:total} for additional constraints). Similar measurements could be made for nearby barred galaxies (see \S\ref{sec:extra}), paying particular attention to spectral features that signify a starburst 50--100~Myr ago with no recent star formation, such as strong Balmer absorption lines and weak nebular emission lines \citep[which is relatively straightforward using fiber spectra from the Sloan Digital Sky Survey, see][]{wild07}. If such a long period were indeed found, it would indicate that the slow evolution of the gas to becoming gravitationally unstable may not be the key mechanism for driving the episodicity, but instead the variation of the gas inflow rate along the bar could be important.

\subsection{Implications and tests of suppression by turbulence} \label{sec:testturb}
In \S\ref{sec:turb}, we show that a very promising explanation for the low SFR in the CMZ is an elevated volume density threshold for star formation due to the high turbulent pressure (with ${\cal M}\sim30$), based on the star formation models of \citet{krumholz05} and \citet{padoan11}. This provides a theoretical justification for the conclusion made in \S\ref{sec:obs} that the observed SFR is inconsistent with the universal threshold density from \citet{lada12}.

Even though the CMZ greatly extends the range of environmental conditions that can be probed in the Milky Way, a conclusive verification of the turbulence hypothesis requires our analysis to be extended to external galaxies. If the volume density threshold for star formation is as variable as would be expected on theoretical grounds, then there should also be an environmental variation of the molecular line transitions that correlate with star formation tracers. In high-pressure environments, transitions with low critical densities such as CO(1--0) should no longer correlate tightly with star formation tracers. Instead, molecules with high-critical density transitions such as HCN ($n=10^5$--$10^6~{\rm cm}^{-3}$) or vibrationally excited transitions should take over. Before the arrival of the Atacama Large Millimeter Array (ALMA), the observational issue was that transitions with high critical densities required unfeasibly high-sensitivity observations. However, in the next couple of years it should be possible to test the spatial correlation between star formation and various dense gas tracers at high spatial resolution using ALMA.

Within our own Galaxy, ALMA observations of clouds in the CMZ are extremely well-suited for addressing the variation of the threshold density for star formation. At the distance of the CMZ, sub-arcsecond scales need to be probed in order to resolve the final stages of the cloud fragmentation and collapse towards cores and protostars. At the distance of the CMZ, only interferometers like ALMA are capable of reaching the necessary sensitivity and resolution in the wavelength range of interest. By combining star formation tracers such as cm continuum with high-density gas tracers, it will be possible to map the conversion of gas into stars in detail. A comparison with similar data from the solar neighbourhood should lead to a conclusive picture of the environmental variation of density thresholds for star formation.

A final experiment for testing if the variation of the critical density threshold for star formation explains the low SFR in the CMZ would be to constrain the density PDF in the CMZ and compare its shape to the theoretically expected form. Ideally, the PDF would need to be measured well enough to verify the possible influence of weak magnetic fields (see \S\ref{sec:localobs}). It is possible to map the PDF by comparing the total flux above certain molecular line transitions. The one caveat is that these transitions will need to probe densities as high as the suggested threshold density of $n_{\rm th}\ga3\times10^7~{\rm cm}^{-3}$.

It is also relevant to establish whether the IMF in the CMZ may be bottom-heavy due to the high Mach number and correspondingly low sonic mass. Current IMF determinations are only capable of reaching masses of $m\geq5~\msun$ \citep[e.g.][]{bastian10,hussman12}, but with the new generation of large-scale facilities like the E-ELT it should be possible to probe the IMF at $m<1~\msun$. A presently possible, but less direct method is to perform ALMA observations of the CMF in star-forming clouds in the CMZ. While this assumes a certain mapping of the CMF to an IMF, the advantage is that this provides a more direct test of star formation theories employing the sonic mass or thermal Jeans mass. These theories actually predict a CMF rather than an IMF and hence a measurement of the CMF would put direct constraints on the physics of star formation in the CMZ.

\section{Discussion} \label{sec:disc}

\subsection{Summary} \label{sec:summ}
We find that processes on both global and local scales may play a role in the reduction of the star formation rate in the Central Molecular Zone of the Milky Way with respect to density-dependent ($N>1$ in equation~\ref{eq:sflaw}) star formation relations.
\begin{enumerate}
\item
The gas by itself is not strongly self-gravitating and is held together by the stellar potential, which may slow the rate at which gas clouds can become gravitationally unstable. (\S\ref{sec:surfacethr})
\item
Star formation could be episodic due to a gradual build-up of dense gas by spiral instabilities and its rapid consumption once the density threshold for gravitational instability is reached. Alternatively, variations in the gas inflow rate along the Galactic bar could play a role. In both cases, the present state of the CMZ corresponds to a near-minimum in the SFR. (\S\ref{sec:episodic})
\item
Crucially, the high turbulent pressure in the CMZ increases the volume density threshold for star formation to $n_{\rm crit}\ga3\times10^7~{\rm cm}^{-3}$, which is orders of magnitude beyond the \citet{lada10} threshold derived empirically for the solar neighbourhood, and hence substantially decreases the SFR with respect to such an empirical threshold. (\S\ref{sec:turb}) To first order, this explains why the CMZ \citep{longmore13} and many of its constituent clouds \citep{longmore12,kauffmann13} are presently not efficiently forming stars. We discuss the possible mechanisms for driving the turbulence below.
\end{enumerate}
Interestingly, the volume density threshold implied by turbulent pressure for solar-neighbourhood conditions is $n\sim10^4~{\rm cm}^{-3}$, which coincides with the \citet{lada10} threshold (see \S\ref{sec:localobs} and \S\ref{sec:turb}). This agreement theoretically supports the idea that a universal, fixed density threshold for star formation is ruled out by the observations, and that the empirical threshold reflects the solar-neighbourhood value of an environmentally dependent threshold.

A wide range of mechanisms is found to be unable to explain the dearth of star formation in the CMZ.
\begin{enumerate}
\item
When including the stellar potential, the gas disc is found to be only marginally Toomre stable, and hence susceptible to gravitational collapse. While this rules out morphological quenching, note that the gas by itself is not strongly self-gravitating, which {as stated previously may slow the rate at which gas clouds can become gravitationally unstable} and thereby affect the SFR at a given gas density. (\S\ref{sec:surfacethr})
\item
Cloud lifetimes are not limited by dynamical heating due to tidal interactions. In principle this could lead to lower SFEs, but the tidal perturbations necessary to heat the clouds dynamically also induce tidal compression and therefore enhance the energy dissipation rate. (\S\ref{sec:shocks})
\item
The tidal density is lower than the mean volume density of the molecular gas in the CMZ, indicating that star formation is not inhibited due to clouds being tidally stripped. (\S\ref{sec:tidal})
\item
An underproduction of massive stars due to a possibly bottom-heavy IMF only introduces a factor of $\sim2$ uncertainty in the measured SFR, much less than the observed discrepancy. (\S\ref{sec:imf})
\item
The atomic-to-molecular phase transition of hydrogen occurs at too low a surface density to play a role in the CMZ, especially at the high metallicity of the bulge. (\S\ref{sec:phase})
\item
Given the current constraints on the magnetic field strength, we find that the magnetic field is capable of affecting the density PDF of the ISM somewhat (\S\ref{sec:localobs}). However, the magnetic pressure in the CMZ is much lower than the turbulent pressure. (\S\ref{sec:magnetic})
\item
Radiation pressure should dominate the feedback energy at the surface density of the CMZ, but a SFE larger than 100~per~cent would be necessary to have radiation pressure compete with the turbulence throughout the CMZ. (\S\ref{sec:rad})
\item
Cosmic rays, produced in supernovae or by the central black hole, are unable to overcome the turbulent pressure, even if the SFR was several orders of magnitude higher in the past. By contrast, they do contribute to the thermal pressure in the 100-pc ring. (\S\ref{sec:rays})
\end{enumerate}
While each of the above effects does not affect the SFR sufficiently to explain the low SFR in the CMZ, they could work in conjunction with a high volume density threshold due to turbulence. The conclusion that feedback processes are not solely responsible for the low SFR in the CMZ is also supported by the discrete distribution of the main star formation events -- if feedback did play a role, it would only be able to do so locally.

It is interesting that the low SFR in the CMZ only agrees with one global star formation relation -- it provides an almost perfect extrapolation of the \citet{bigiel08} relation (\S\ref{sec:globalobs}). Is this the `fundamental' star formation relation? The mere existence of high-redshift starburst galaxies with $\Sigma_{\rm SFR}$ two orders of magnitude higher than the CMZ at the same gas surface density \citep{daddi10b} suggests otherwise. As discussed throughout this paper, there are several reasons why the CMZ could be abnormal, and the extrapolation of the \citet{bigiel08} relation to the CMZ therefore seems premature. The observational and numerical tests of the proposed mechanisms of \S\ref{sec:impl} will be essential to draw further conclusions on this matter.

In the remainder of this paper, we use the above findings to construct a self-consistent picture of star formation in the CMZ, as well as its implications for star formation in the centres of other galaxies.

\subsection{What is driving the turbulence?} \label{sec:driver}
We find that (a combination of) more than one mechanism can be tuned to reach a satisfactory estimate of the SFR in the CMZ. However, it remains to be determined which mechanism is actually responsible for inhibiting star formation.

Out of the possible explanations listed in \S\ref{sec:summ}, the high turbulent pressure of the CMZ is perhaps the most fundamental. Contrary to the possible episodicity of the SFR, or perturbations acting on the cloud scale, it is the only mechanism that works instantaneously, at this moment in time. However, attributing the low SFR to turbulence alone is an incomplete argument, because the turbulent energy dissipates on a vertical disc crossing time. {Using the numbers from Table~\ref{tab:prop}, the dissipation rate in the 230-pc integrated CMZ is roughly $4\times10^{-22}~{\rm erg}~{\rm cm}^{-3}~{\rm s}^{-1}$ \citep{maclow04}. The dissipation time-scale} is a mere $t_{\rm diss}=2h/\sigma\sim7$~Myr, whereas for the 100-pc ring it is about $1$~Myr. Something is needed to maintain the turbulence.

\subsubsection{Gas inflow along the Galactic bar} \label{sec:inflow}
Could the gas inflow along the bar be responsible for driving the turbulence? If we assume a steady-state CMZ, then it is straightforward to calculate the kinetic energy of the gas inflow along the bar. The energy flux of the inflow is given by $v_{\rm infl}^2/t_{\rm infl}$, where $v_{\rm infl}$ is the gas inflow speed, and $t_{\rm infl}\equiv M_{\rm gas}/\dot{M}_{\rm infl}$ is the time-scale for accumulating the present gas mass. The energy dissipation rate is $\sigma^2/t_{\rm diss}$. Energy balance then gives
\begin{equation}
\label{eq:vinfl}
v_{\rm infl}=\sigma\sqrt{\frac{t_{\rm infl}}{t_{\rm diss}}} .
\end{equation}
In a steady-state CMZ, the gas inflow time-scale must equal the time-scale for gas depletion $t_{\rm depl}$, which in the CMZ is about a Gyr (see Table~\ref{tab:prop}). Substituting values representative of the CMZ yields a required $v_{\rm infl}$ for maintaining the turbulence of $v_{\rm infl}\sim200~{\rm km}~{\rm s}^{-1}$, which is comparable to the highest line-of-sight velocity $v_{\rm los}^{\rm max}\sim200~{\rm km}~{\rm s}^{-1}$ that can be seen in the HOPS NH$_3(1,1)$ data. However, this is likely not the radial velocity component -- given the observed phase-space distribution of the gas, a more representative inflow speed along the bar would be $v_{\rm infl}\la100~{\rm km}~{\rm s}^{-1}$. Due to line-of-sight confusion this is a tentative estimate -- detailed modelling is necessary to derive the gas inflow speed with satisfactory accuracy.

If we relax the condition of a steady-state CMZ, a gas inflow time-scale of $t_{\rm infl}\sim300~{\rm Myr}$ (rather than the adopted $t_{\rm infl}=1~{\rm Gyr}$) would be needed to maintain the current turbulent pressure at an inflow speed of $v_{\rm infl}\sim100~{\rm km}~{\rm s}^{-1}$. This time-scale lies within the observed constraints on the mass deposition rate, even when accounting for the removal of mass by outflows \citep[$t_{\rm infl}=60$--$500$~Myr, see][]{crocker12b}. In order for this scenario to be viable, the peak SFR would have to be roughly five times higher than is presently observed. This would put the CMZ temporarily on the star formation relations of equations~(\ref{eq:sflaw}) and~(\ref{eq:sflawomega}).

We find that the gas inflow may be capable of driving the turbulence, but only if star formation in the CMZ is episodic, indicating that the current gas depletion time-scale is not an accurate estimate for the inflow time-scale. {There is no unique combination of gas inflow and star formation histories that explains the current degree of turbulence as the result of the global gas inflow.} For instance, we might be seeing the outcome of a persistently high inflow rate, which builds up gas until finally a starburst is generated -- with the CMZ presently at the stage between two bursts. Alternatively, a low gas inflow rate would not be able to drive all of the turbulence, but could still substantially increase the time-scale for the energy dissipation.

In summary, the gas inflow along the Galactic bar is a promising turbulence driver if the SFR in the CMZ is episodic. Whether or not this mechanism can explain the observed turbulence fully is uncertain due to the observational difficulty of characterising the inflow. We are currently modelling the gas inflow in more detail, the results of which will be presented in a future paper (Krumholz \& Kruijssen, in preparation).

\subsubsection{Feedback ejecta} \label{sec:ejecta}
Episodic starbursts would periodically drive superbubbles into the low-density ISM. As mentioned in \S\ref{sec:rays}, massive star feedback can locally blow out parts of the 100-pc ring, producing the observed asymmetries in CO, NH$_3$ and cold dust (see Figure~\ref{fig:img}). Radio and $\gamma$-ray observations of nested giant bubbles indeed provide evidence that stellar feedback has a profound impact on localized portions of the CMZ \citep{sofue84,su10,carretti13}. On large scales, the bubbles consist of swept up, low-density gas, whereas higher density material can also be driven to a bubble, but only on size scales $\Delta R \la h$.

After cooling in the Galactic halo, these bubbles return to the disc as high-velocity clouds \citep[e.g.][]{bregman80,wakker97}. Provided that the resulting energy flow is sufficiently high, this may drive the turbulence {in the {\it dense} ISM}. However, we estimate in \S\ref{sec:rays} that even at a star formation maximum, the energy density of feedback is probably smaller than the turbulent pressure. After conservation of energy (or a decrease due to dissipative losses) this implies that in-falling clouds are not likely to provide the necessary energy.

Even if we imagine that feedback would be the source of the turbulence, then the present low SFR and lack of energy input by feedback implies that star formation would start after the turbulent energy has dissipated. This would lead to prevalent star formation throughout the CMZ in a few Myr from now. The resulting feedback would compress gas, which could trigger even more star formation. Only if the gas were converted to a warm, diffuse phase, star formation could in principle be halted by feedback. However, the resulting blow-out would channel away the energy and a molecular disc would remain, which should still be forming stars.

Quantifying the effects of feedback and other classical turbulence drivers shows that neither are effective. Following the compilation by \citet{maclow04}, we find that magnetorotational instabilities, gravitational instabilities, protostellar outflows, ionizing radiation, and supernovae yield heating rates of $\{6\times10^{-24},1\times10^{-24},5\times10^{-26},2\times10^{-23},3\times10^{-23}\}~{\rm erg}~{\rm cm}^{-3}~{\rm s}^{-1}$, respectively. Hence, they all fall short of compensating the dissipation rate ($4\times10^{-22}~{\rm erg}~{\rm cm}^{-3}~{\rm s}^{-1}$) by at least an order of magnitude. Only within the 100-pc ring, the heating rate due to gravitational instabilities ($1\times10^{-21}~{\rm erg}~{\rm cm}^{-3}~{\rm s}^{-1}$) is similar to the local turbulent energy dissipation rate ($2\times10^{-21}~{\rm erg}~{\rm cm}^{-3}~{\rm s}^{-1}$). {Hence, during the gravitational collapse of clouds like the Brick, their turbulent energy dissipation time-scale can be extended by up to a factor of a few.}

\subsubsection{Secular instabilities}
In conjunction with inflow-driven turbulence, another sustainable solution would be that the turbulence is driven by (acoustic) gas instabilities themselves. {In the Galactic disc, spiral waves are driven by gravitational instabilities, but this process fails within the inner Lindblad resonance of a disc galaxy, where the waves become pressure-driven \citep{montenegro99}. As non-self-gravitating gas falls into the stellar mass-dominated potential of the CMZ, the geometric convergence compresses the gas to a higher density, even in the absence of self-gravity. This acoustic instability drives a spiral wave, in which the turbulent pressure increases \citep{dobbs11b}. When the gas density has built up far enough that the compression increases the gas density beyond the stellar density and hence leads the gas to become self-gravitating, then the rate of energy dissipation increases and the gas may eventually collapse and form stars.}\footnote{Note that the regions that reach the threshold density for gravitational collapse need not be randomly distributed throughout the CMZ, but may be corresponding to specific locations in the geometry, such as possibly Sgr~B2 (see \S\ref{sec:shocks} and \citealt{elmegreen94,boeker08,sandstrom10}).}

In this picture, the inner gas discs of galaxies are driven to a spiral catastrophe by acoustic instabilities, which drive the turbulence and produce irregular gas structures\footnote{There is observational evidence for the presence of such dust lane structures in the central region of the Milky Way on larger ($\sim~{\rm kpc}$) scales (see e.g. \citealt{mccluregriffiths12} and references therein).} during an intermittent phase in which a minor role is played by the self-gravity of the gas and star formation. This scenario is consistent with the idea that the central rings of galaxies are only prone to gravitational instabilities when a certain density threshold is reached (see \S\ref{sec:episodic}). For this scenario to be viable, the gas inflow rate needs to be small enough to ensure that the density threshold for gravitational instability is not always satisfied -- otherwise, acoustic instabilities would not be able to drive the observed turbulence and star formation would no longer be (periodically) suppressed. In \S\ref{sec:conditions}, we discuss a simple empirical criterion to determine in which systems star formation is episodic.

\subsection{A self-consistent star formation cycle} \label{sec:total}
\begin{figure*}
\center\resizebox{\hsize}{!}{\includegraphics{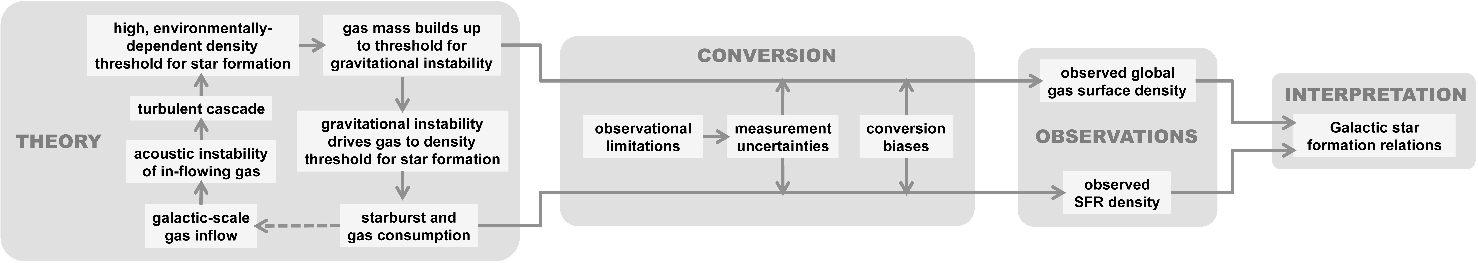}}\\
\caption[]{\label{fig:schem}
      {Schematic representation of a possible, unified picture of local and global processes that control star formation in the CMZ, as well as the conversion to observables and galactic star formation relations. {The bottom arrow in the theory block is dashed because it only indicates the progression of time -- the starburst and gas consumption do not affect the galactic-scale gas inflow.} See \S\ref{sec:total} for a more detailed discussion.}
                 }
\end{figure*}
We have now discussed all the ingredients necessary to construct a single, self-consistent scenario for star formation in the CMZ. In this scenario, the local and global explanations for the low SFR in the CMZ represent different aspects of the same mechanism. Such a holistic approach is crucial for explaining the low SFR in the CMZ, because the different size scales influence each other. For instance, the overall gas supply is regulated on a global scale ($\Delta R>h$), but local physics must determine the gas consumption rate if turbulence sets a critical density for star formation.

\subsubsection{Schematic representation of the cycle}
Based on the discussion in \S\ref{sec:summ} and \S\ref{sec:driver}, we propose a multi-scale cycle that controls the star formation in the CMZ. It is summarized in Figure~\ref{fig:schem}, which also illustrates the conversion of physical quantities to the observables that part of the analysis in this paper is based on. The different stages of the cycle are as follows.
\begin{enumerate}
\item \label{pt:inflow}
Gas flows towards the CMZ from larger galactocentric radii, which can be driven by secular evolution (e.g.~dust lane or bar transport) or by external torques (e.g.~due to minor mergers or galaxy interactions). The gas is not self-gravitating and is held together by the gravitational potential of the stars, of which the volume density exceeds that of the gas.
\item
Geometric convergence causes the inflowing gas to be compressed by acoustic instabilities, which drives pressure waves even when the gas is not gravitationally unstable.
\item
The combination of the global gas inflow and acoustic instabilities drive the highly-supersonic turbulence in the CMZ, which cascades down to smaller scales.
\item \label{pt:turb}
The elevated turbulent pressure increases the local volume density threshold for star formation as $n_{\rm th}\propto n_0{\cal M}^2$ to $n_{\rm th}\ga 3\times10^7~{\rm cm}^{-3}$ (this is a lower limit due to the ill-constrained virial ratios of the clouds).
\item \label{pt:accum}
Due to the high density threshold and the lack of gravitational collapse, the gas is not consumed to form stars and instead accumulates until the gas density becomes comparable to the stellar density, when it finally becomes self-gravitating and susceptible to gravitational instabilities.
\item
The gravitational instabilities allow the gas density PDF to develop a power-law tail, driving the gas to sufficiently high densities to overcome the local density threshold for star formation.
\item \label{pt:burst}
At such high densities, the time-scale for turbulent energy dissipation is short, and the gas is rapidly turned into stars. This can occur at different times in different parts of the CMZ, as is exemplified by Sgr~B2 undergoing a starburst while the other CMZ clouds are inactive.
\item
The cycle repeats itself, starting again from point (i).
\end{enumerate}
Note that in this cycle, the rate-limiting factor is the slow evolution of the gas towards gravitational collapse. Once star formation is initiated, it should proceed at the rapid pace appropriate for the high densities involved (see below). Because the gas needs to reach high densities before becoming gravitationally unstable and forming stars, it is likely that much of the star formation in the CMZ occurs in bound stellar clusters \citep[see \S\ref{sec:episodic} and][]{kruijssen12d,longmore14}.

\subsubsection{Relevant time-scales} \label{sec:time}
Whether or not the system is observed during a starburst or a star formation minimum depends on the relative time-scales of the stages in the above cycle. The inflow and accumulation of gas from stages~\ref{pt:inflow}--\ref{pt:turb} take place on a dynamical time-scale, which is 5--10~Myr on the spatial scale of the 100-pc ring, and $\sim100$~Myr at the outer edge of the bar. These widely different time-scales are applicable under different circumstances. In the special case of a CMZ-wide starburst, stages~\ref{pt:inflow}--\ref{pt:turb} would be governed by global gas dynamics and would last a gas inflow time-scale of $t_{\rm infl}\sim300$~Myr (see \S\ref{sec:inflow}). However, if the CMZ undergoes multiple, localized starbursts, then the gas should be resupplied to the starburst region(s) on the dynamical time-scale of the CMZ and hence stages~\ref{pt:inflow}--\ref{pt:turb} take place on a $\sim10$~Myr time-scale.

As stated in \S\ref{sec:surfacethr}, the time-scale for clouds to become gravitationally unstable during stage~\ref{pt:accum} is $t_{\rm grav}\sim Q_{\rm gas}/\kappa$ \citep[e.g.][]{jogee05}, which in the 100-pc ring is $t_{\rm grav}\sim1~{\rm Myr}$, but in the 230~pc-integrated CMZ it is $t_{\rm grav}\sim20~{\rm Myr}$. This comparison shows that the 100-pc ring may be the unstable phase of the proposed cycle, and that stage~\ref{pt:burst} may be reached at different times throughout the CMZ.

On the scale of individual clouds, the gas consumption time-scale during stage~\ref{pt:burst} should be shorter than the time spanned by stages~\ref{pt:inflow}--\ref{pt:accum}, because the free-fall time is only a few $10^5$~yr or less at the high density of the clouds in the CMZ. However, the integrated starburst duration for the entire CMZ depends on mixing and hence should be comparable to its dynamical time-scale (i.e.~up to 10~Myr). Such relatively brief episodes of nuclear activity are supported by Hubble Space Telescope (HST) observations of nearby galaxies \citep{martini03}.

We are thus left with a simple picture in which the CMZ underproduces stars for 10~Myr or 100~Myr (depending on whether the starbursts are localized or CMZ-wide), and then consumes its gas over a 10-Myr time-scale. In the localized-starburst case, it is therefore equally likely to observe the CMZ at its current, low SFR as it is to observe it at a maximum. In the global-starburst case, the CMZ is most likely to be observed during its inactive phase. This simple picture may be confused by the simultaneous occurrence of mini-bursts of star formation in different parts of the CMZ, which could temporarily place the region on or above galactic star formation relations, even during a global minimum.

\subsubsection{Comparison to galactic star formation relations}
We now compare the CMZ to galactic star formation relations as a function of spatial scale to distinguish between the above localized-starburst and global-starburst scenarios. While it is tempting to go one step further and modify galactic star formation relations to accommodate our proposed scenario in its entirety, this is not feasible because the position in the $\{\Sigma,\Sigma_{\rm SFR}\}$ plane of the CMZ and its constituent parts strongly evolves with time. Such a modification would require quantifying the cycle of Figure~\ref{fig:schem} through space and time from the single snapshot we can presently observe.

The conversion of the proposed cycle to observables is well beyond the scope of this paper, but Figure~\ref{fig:schem} shows a qualitative illustration. The global gas surface density mainly samples stage~\ref{pt:accum}, whereas the star formation rate surface density traces stage~\ref{pt:burst} of the cycle. During these stages, the CMZ should lie below and above the galactic star formation relations, respectively, by an amount that depends on the detailed physics of the system.

As illustrated in Figure~\ref{fig:schem}, observations of the gas and SFR surface densities very indirectly probe the physics of star formation, because their relation to the actual volume densities of star-forming gas and young stars depends on several factors. Observational limitations such as the spatial resolution, the sensitivity, and possible tracer biases lead to measurement uncertainties in the derivation of the relevant physical quantities. For instance, unresolved stars and gas may not be occupying the same volume \citep[e.g. due to different filling factors or porosities,][]{silk97}, and the sampling of the gas (PDF) and the star formation may be incomplete (the degree of which depends on the adopted tracers). Notwithstanding these complications, we attempt a comparison of the CMZ to galactic star formation relations to characterise the cyclic processes that govern the evolution of the CMZ.

One quantitative constraint follows from Figure~\ref{fig:kslaw}, which shows that on a certain length scale ($\Delta R=230~{\rm pc}$) the CMZ fits the empirical Schmidt-Kennicutt relation of equation~(\ref{eq:sflaw}), although it does not satisfy the Silk-Elmegreen relation of equation~(\ref{eq:sflawomega}). \citet{krumholz12a} recently proposed that a universal star formation relation is obtained by modifying the Silk-Elmegreen relation. They substitute $\Omega\rightarrow1/t_{\rm sf}$ in equation~(\ref{eq:sflawomega}), where the star formation time-scale $t_{\rm sf}$ is either set by the local (GMC) or global (galactic) dynamical time-scale, whichever is the shortest. If we account for the long condensation time-scale of GMCs and hence assume that $t_{\rm sf}\sim t_{\rm grav}\sim Q_{\rm gas}/\kappa$ (as in the Toomre regime of \citealt{krumholz12a}), we see that the 230~pc-integrated CMZ fits a modified Silk-Elmegreen relation:
\begin{equation}
\label{eq:sflawomegamod}
\Sigma_{\rm SFR}=A_{\rm SE}\Sigma/t_{\rm grav}\sim A_{\rm SE}\Sigma\Omega/Q_{\rm gas} .
\end{equation}
This modification still does not agree with the 1.3$^\circ$ complex and the 100-pc ring onto the relation, because they have lower $Q_{\rm gas}$ than the 230~pc-integrated CMZ.

It is intruiging that this simple self-gravity argument reconciles the large-scale properties of the CMZ with the Silk-Elmegreen relation, whereas it still fails on smaller scales. In \S\ref{sec:episodic}, we considered the option that on a 230-pc scale we might sample all of the evolutionary stages of the cloud-scale star formation process, causing the integrated CMZ to agree with the Schmidt-Kennicutt relation. This explanation was discarded, because adequate statistical sampling should already be achieved above 80-pc scales and hence the 100-pc ring should have been consistent with the same relation. However, this argument required that the clouds independently occupy different evolutionary stages.

The plausible existence of a large-scale episodic cycle now shows that the star formation activity of clouds in the CMZ likely {\it is} synchronised, i.e.~clouds evolve collectively from being quiescent to actively forming stars. The key question is on which spatial scale they are synchronised. If the CMZ undergoes starbursts in its entirety,  the evolutionary stages of the clouds are linked across the entire region. In the framework of \citet{kruijssen14}, such an interdependence of cloud evolution is synonymous to stating that the CMZ represents only a single `independent region', implying that the CMZ should not obey galactic star formation relations during most of its evolutionary cycle, just like a single cloud does not fit galactic star formation relations \citep[e.g.][]{heiderman10}. If the CMZ experiences smaller-scale, localized starbursts that cover multiple clouds, it may still consist of several independent regions -- possibly sufficient to warrant the adequate statistical sampling of the time sequence of star formation.

The agreement of the 230~pc-integrated CMZ with the Schmidt-Kennicutt relation and the modified Silk-Elmegreen relation of \citet{krumholz12a} suggests that there has been no recent CMZ-wide starburst. Instead, the CMZ likely undergoes multiple, localized starbursts. This resolves the ambiguity we were left with in \S\ref{sec:time} and allows us to quantify the time-scales involved in the proposed cycle. (Parts of) the CMZ underproduce(s) stars for 10~Myr and then consume(s) its gas over a similar time-scale. We therefore predict that in other galaxies similar to the Milky Way, the part of the population with enhanced nuclear SFRs should be comparable to those with suppressed nuclear SFRs.

\subsubsection{Generalisation to large-scale boundary conditions} \label{sec:conditions}
The cycle of Figure~\ref{fig:schem} heavily relies on the gradual build-up of gas up to a certain threshold density above which it collapses. However, this does not always occur. There exists a global gas convergence rate above which the threshold density criterion for gravitational instability is always satisfied and star formation is no longer episodic. This may be one of the differences between the CMZ and the vigorously star-forming galaxies of \citet{daddi10b}, which are thought to be mergers undergoing large-scale mass convergence towards a common point of reference \citep{genzel10}.

The precise value of a critical convergence rate above which gas rapidly becomes gravitationally unstable varies with galaxy properties, but empirically it is straightforward to distinguish between three gas convergence regimes.
\begin{enumerate}
\item
If a system is in a pre-starburst phase and the gas mass is building up, then the convergence rate $\dot{M}_{\rm in}$ must exceed the sum of the SFR and the mass outflow rate due to feedback \citep[which can be non-negligible, see][]{bolatto13}, even if the convergence rate is variable.
\item
If the convergence rate is so high that the gas mass does not have to build up before it becomes gravitationally unstable and forms stars, the SFR is comparable to the convergence rate. This should also hold instantaneously and would therefore be insensitive to variations of the convergence rate, as long as the convergence rate is high enough to persistently drive star formation.
\item
If a system is undergoing a starburst, the SFR should exceed the convergence rate by definition, and hence the mass consumption rate including feedback $\dot{M}_{\rm out}$ does so too.
\end{enumerate}

Summarising the above three regimes, we predict that a roughly constant SFR due to persistent, convergence-driven gravitational instabilities requires a mass convergence rate $\dot{M}_{\rm in}\sim\dot{M}_{\rm out}$, whereas episodic star formation takes place whenever $\dot{M}_{\rm in}\neq\dot{M}_{\rm out}$. This balance between the inflow and outflow rates can be evaluated empirically. Because it contains the physics of star formation and feedback, the distinction between the three regimes should hold universally. Galaxies (or their centres) with $\dot{M}_{\rm in}<\dot{M}_{\rm out}$ should follow the \citet{daddi10b} sequence of galaxies with enhanced SFRs, whereas systems with $\dot{M}_{\rm in}>\dot{M}_{\rm out}$ should statistically be found to have lower SFRs (making excursions to elevated SFRs and hence $\dot{M}_{\rm in}<\dot{M}_{\rm out}$ during their periodic bursts of star formation).

\subsection{Implications for (extra)galactic nuclei} \label{sec:extra}
Given that the CMZ can only be observed in its present state, the cycle of Figure~\ref{fig:schem} is sampled better by including other galaxies where the same process is taking place. The inner regions of spiral galaxies often host rich dust structures in which the number of spiral arms increases with radius, with power spectra that are consistent with the presence of acoustic instabilities \citep[e.g.][]{elmegreen98,martini99,elmegreen02b}. These structures host little to no star formation, which if present is often arranged in a circumnuclear ring \citep[e.g.][]{barth95,jogee02,sandstrom10,vanderlaan13}, and their gas by itself is often not strongly self-gravitating \citep{sani12}.

The physical conditions in the central regions of these galaxies are almost indistinguishable from those in the CMZ and hence the star formation histories may be similar too. Indeed, there are several nearby galaxies of which the star formation activity in their central regions falls below galactic scaling relations \citep{hsieh11,nesvadba11,sani12,saintonge12}, even in nuclei that at first sight appear to be undergoing a starburst \citep{kenney93}. By contrast, examples of galaxy centres with a normal or enhanced SFR exist as well, which may indicate a different stage of the same process (e.g.~the central starbursts in the sample of \citealt{jogee05}, also see \citealt{sakamoto07,sakamoto11,leroy13}). {Finally, it has been suggested that the SFR in galaxy centres is consistent with galactic scaling relations \citep{fischer13}.

The wide range of different SFRs observed in galactic nuclei could potentially be consistent with an episodic cycle as proposed in this paper. However, it is crucial to ensure that any comparison between the CMZ and other galactic nuclei considers similar size scales and spatial resolutions. The extragalactic observations} often evaluate star formation relations by smoothing over scales of 0.5--1~kpc, on which the CMZ is actually consistent with the star formation relations of disc galaxies. Only when probing the detailed surface or volume densities of the gas, deviations begin to emerge.

We suggest that a high-resolution survey of the star formation activity in galactic nuclei is necessary to quantify the duration of the different stages in Figure~\ref{fig:schem} across the galaxy population. This is a key step, which is yet to be made for a large galaxy sample. \citet{martini03} considered HST/WFPC2 observations of a sample of 123 nearby, active and passive, barred and unbarred galaxies. While these authors do note that their sample was not designed to study star formation in detail, they use dynamical arguments to infer that the nuclear star formation episode may last up to several Myr, which is consistent with what we propose for the CMZ in \S\ref{sec:total}.

Following a similar argument, \citet{davies07} find that active galactic nuclei (AGN) become active some 50--100~Myr after the onset of prior (but ceased) starburst activity. If the starburst occurs throughout the galactic nucleus instead of in a localized fashion (see \S\ref{sec:time}), this can be understood in terms of the scenario of Figure~\ref{fig:schem} -- a nucleus-wide starburst and its corresponding feedback may either consume or eject the gas, preventing the black hole from being fed until supernova feedback has ended (after 40~Myr). After that time, the gas flow towards the black hole continues. In our scenario, that gas is not (yet) self-gravitating, and hence it can proceed to the black hole without forming stars. This phase should be characterized by centrally concentrated stellar surface brightness and gas mass profiles due to the combination of prior starburst activity and a restored gas inflow, as has recently been found by \citet{hicks13}. Hence, the phase of low star formation activity in the cycle of Figure~\ref{fig:schem} could provide a window for the efficient growth of massive black holes.

A key question is how the general the physics are that regulate star formation in the CMZ and the central regions of nearby galaxies. Do they translate easily to other extragalactic environments? Giant elliptical galaxies provide an interesting target, because in recent years they have been found to contain a sizable, but remarkably quiescent gas reservoir \citep[e.g.][]{crocker12}. Another, less obvious comparison should be made between the CMZ and ULIRGs and high-redshift, star-forming galaxies, which in terms of their gas properties are almost indistinguishable \citep{kruijssen13}. The similarity of their gas properties aside, these galaxies are undergoing vigorous starbursts, whereas the SFR in the CMZ is much lower than expected. Some rough arguments for these similarities and differences have been made in this paper, but future work should address the underlying physical reasons in the necessary detail.

In summary, we find that the lower-than-expected SFR in the CMZ can be explained by a self-consistent cycle of star formation activity, which is is presented in \S\ref{sec:total} and Figure~\ref{fig:schem}. It connects the galaxy-scale gas inflow, acoustic and gravitational instabilities, turbulence, and local star formation thresholds. In this cycle, the SFR is episodic. During the quiescent phase, the SFR is limited by the slow evolution of the gas towards collapse -- it continually builds up until a critical density for gravitational instability is reached. During the subsequent active phase, star formation proceeds at a normal rate. We conclude that a low SFR may well be a common property of the centres of Milky Way-like galaxies, although galactic nuclei with elevated SFRs should exist too if our proposed explanation holds. There are tantalising indications that the lessons learned about star formation in the CMZ are relevant to other (extra)galactic environments, from galactic nuclei in general to giant elliptical galaxies and high-redshift starbursts. These are exciting avenues for future work aiming to characterise galactic-scale star formation in extreme environments.

\section*{Acknowledgments}
We thank the referee, Eric Keto, for helpful suggestions that greatly improved the presentation of this work. JMDK thanks Eli Bressert, Andreas Burkert, Cathie Clarke, Richard Davies, Maud Galametz, Phil Hopkins, Mark Morris, Thorsten Naab, Stefanie Walch, Simon White, Farhad Yusef-Zadeh, and Tim de Zeeuw for helpful discussions. We are indebted to Timothy Davis and Maud Galametz for detailed comments on the manuscript. JMDK thanks the Institute of Astronomy in Cambridge, where a large part of this work took place, for their kind hospitality and gratefully acknowledges support in the form of a Visitor Grant. JMDK, SNL, NM and JB acknowledge the hospitality of the Aspen Center for Physics, which is supported by the National Science Foundation Grant No.~PHY-1066293.

This research made use of data products from the Midcourse Space Experiment. Processing of the data was funded by the Ballistic Missile Defense Organization with additional support from NASA Office of Space Science. This research has also made use of the NASA/IPAC Infrared Science Archive, which is operated by the Jet Propulsion Laboratory, California Institute of Technology, under contract with the National Aeronautics and Space Administration.

\bibliographystyle{mn2e}
\bibliography{mybib}

\bsp

\label{lastpage}

\end{document}